\documentclass[oldversion]{aa}
\usepackage{epsfig,times}
\usepackage{amssymb}
\usepackage{txfonts}
\usepackage{graphicx}
\usepackage{natbib}
\def \be{\begin{equation}}
\def \ee{\end{equation}}
\def \beq{\begin{eqnarray}}
\def \eeq{\end{eqnarray}}

\newcommand{\rxa}{\hbox{RX\,J0420.0$-$5022}}
\newcommand{\rxb}{\hbox{RX\,J0720.4$-$3125}}
\newcommand{\rxc}{\hbox{RX\,J0806.4$-$4123}}
\newcommand{\rxd}{\hbox{1RXS\,J130848.6+212708}}
\newcommand{\rxe}{\hbox{RX\,J1605.3+3249}}
\newcommand{\rxf}{\hbox{RX\,J1856.4$-$3754}}
\newcommand{\rxg}{\hbox{1RXS\,J214303.7+065419}}
\newcommand{\rbsd}{\hbox{RBS1223}}
\newcommand{\rbsg}{\hbox{RBS1774}}

\newcommand{\pdot}{\dot P}

\begin{document}

\title{Anisotropic thermal emission from magnetized neutron stars}
\author{J.F. P\'erez--Azor\'{\i}n \and J.A.~Miralles 
\and J.A.~Pons }
%\offprints{}
\institute{Departament de F\'{\i}sica Aplicada, Universitat d'Alacant,
           Ap. Correus 99, 03080 Alacant, Spain}

\date{Received...../ Accepted.....}

\abstract
{The thermal emission from isolated neutron stars is not well understood.
The X-ray spectrum is very close to a blackbody but there is a systematic optical excess
flux with respect to the extrapolation to low energy of the best blackbody fit.
This fact, in combination with the observed pulsations in the X-ray flux, can
be explained by anisotropies in the surface temperature distribution.
We study the thermal emission from neutron stars with strong magnetic fields
$B \ge 10^{13}$ G in order to explain the origin of the anisotropy.
We find (numerically) stationary solutions in axial symmetry of the heat transport
equations in the neutron star crust and the condensed envelope.
The anisotropy in the conductivity tensor is included consistently.
The presence of magnetic fields of the expected strength leads to anisotropy in the
surface temperature. Models with toroidal components similar to or larger than the poloidal field
reproduce qualitatively the observed spectral properties and variability of isolated neutron
stars. Our models also predict spectral features at energies between 0.2 and 0.6 keV
for $B = 10^{13} - 10^{14}$.

}
\titlerunning{Anisotropic thermal emission from magnetized NS}
\authorrunning{P\'erez--Azor\'{\i}n, Miralles \& Pons}

\maketitle

%%%%%%%%%%%%%%%%%%%%%%%%%%%%%%%%%%%%%%%%%%%%%%%%%%%%%%%%%%%%%%%%%%%%%%%%
\section{Introduction}

Neutron stars (NS) with large magnetic fields ($B \geq 10^{13}$ G), the so-called
{\it magnetars}, are becoming more and more abundant as new observations reveal phenomena
that can only be explained by the action of strong magnetic fields. It is now believed
that the small population (4 objects) of soft gamma repeaters (SGRs) are young neutron stars
with magnetic fields in the range $\approx 10^{14}-10^{15}$ G. 
Another subclass of candidates to be magnetars are the anomalous X--ray pulsars (AXPs),
whose high X--ray luminosities and fast spindown rates make them different from isolated 
radio pulsars or from NS in accreting X-ray binaries. The six members of this family
\citep{Tie05,MGa05} exhibit spin periods in the range 5-12 s, and their
inferred magnetic fields (from their period derivative) are in the same range as SGRs
\citep[see e.g.][ for a comprehensive review about these two families of magnetar candidates]{WT05}.

A third rare family of NS, the radio-quiet isolated neutron stars 
among which RX J1856.4-3754 is the first and brightest example 
\citep{Wal96}, shares some common features with the standard magnetars (SGR, AXPs): 
periods clustered in the range 5-10 s., 
and increasing evidence of large magnetic fields ($> 10^{13}$ G). The properties of the
seven confirmed members of this family are summarized in Table \ref{ionstab}.
The most puzzling feature is the apparent optical excess flux (compared to the extrapolation
of the best fit to the X-ray emission) observed in several objects, 
which needs of the existence of large temperature variations over the surface to
reconcile the optical and X-ray spectra \citep{Pons02}.
The evidence of anisotropic temperature distribution is also supported by the fact that
several of the thermal spectra show clear X-ray pulsations with pulsation amplitudes from 5 to
20 \%, while others (RX J1856, RX J1605) have upper limits of 1.3-3\% to the maximum pulsation
amplitude \citep{Bur03,Ker04}, which can be explained in terms of different relative orientations
between the rotation and magnetic field axis.
Thus, it is worth to investigate the influence of strong magnetic field configurations 
on the temperature distribution, which has been shown to be able to
create large anisotropies in neutron star crusts \citep{GKP04},
or in the envelope, where further complications due to 
quantizing effects of the magnetic field or accreted material have been studied in 
detail \citep{Pot03}.
But there is yet another
important issue regarding the thermal emission on magnetized neutron stars.
Below some critical temperature (depending on the composition and
the magnetic field strength), the gaseous layers of highly magnetized
neutron stars may undergo a phase transition that turns the gas into liquid or solid state
\citep{Lai01}, which strongly reduces the emissivity from the NS surface
compared to the blackbody case \citep{Bri80,TZD04,paper1,Lai05}.

In this paper our aim is to extend previous works on the anisotropies and thermal
emission of magnetized neutron stars \citep{GKP04} in two main ways: by extending to 
lower density the calculations, within the model of a condensed surface, and exploring
the effect of toroidal components of the magnetic field.
The generation of toroidal fields
in the early stages of a NS life, and its interplay with the poloidal component is a 
complex problem linked to convective instabilities, 
turbulent mean--field dynamo \citep{BRU03} or the Hall instability
\citep{RKG04}. The magnitude of the toroidal fields is unknown but usually
thought to be larger than the poloidal component and, as we discuss in this paper,
have interesting observational implications.

This paper is organized as follows: in the next section the plasma properties in magnetic
neutron stars are reviewed. In section 3, the magnetic field configurations used in the
calculations are described. In section 4, we describe the equations governing the thermal
evolution and structure in the presence of large magnetic fields, the numerical code used
for the calculations and some tests. The microphysics input is discussed in section 5 and
finally, in section 6, we present our results.

%%%%%%%%%%%%%%%%%%%%%%%%%%%%%%%%%%%%%%%%%%%%%%%%%%%%%%%%%%%%%%%%%%%%%%%%%%%%%%%%%%%%%%%
\begin{table*}
\caption{Properties of isolated neutron stars observed by ROSAT, Chandra, and XMM-Newton 
\cite{Pons02, Hab04, Hab05, KKK03, KK05}.}
\begin{tabular}{lccccccccc}
%\noalign{\smallskip}
\hline
\hline
\noalign{\smallskip}
 Source & kT &  $P$   & $\pdot$ & $\tau$ &  Optical  & Optical & Pulsation & E$_{\rm line}$ & 
B$_{db}$/B$_{cyc}$ \\
        & (eV) &  (s)   & $10^{-12}$ (s~s$^{-1}$) & $10^6$yr &  & excess factor & amplitude
& (keV)  & 10$^{13}$ G \\
\noalign{\smallskip}\hline\noalign{\smallskip}
\rxa        & 45    &  3.453  & $<9$     &         & B$=26.6$  & $<12$ & 0.12    & 0.329   & $<18$/6.6 \\
\rxb        & 85    &  8.391  & $0.07$   & 0.6-2   & B$=26.6$  & 6     & 0.11    & 0.270   & 2.4/5.2 \\
\rxc        & 96    & 11.371  & $<2$     &         & B$>24$    &       & 0.06    & $-$     & $<14$/?  \\
\rxd/\rbsd  & 95    & 10.313  & $<6$     &         & $28.6$    & $<5$  & 0.18    & 0.3     & ?/2-6  \\
\rxe        & 95    &   $-$   & $-$      &         & B$=27.2$  & 11-14 & $<0.03$ & 0.46    & ?/9.5 \\
\rxf        & 60    &   $-$   & $-$      &  0.5    & V$=25.7$  & 5-7   & $<0.02$ & $-$     &  $-$ \\
\rxg/\rbsg  & 101   &  9.437  & $-$      &         & R$>23$    &       & 0.04    & 0.70    & ?/14  \\
\noalign{\smallskip}\hline\noalign{\smallskip}
\end{tabular}
\label{ionstab}
\end{table*}

%%%%%%%%%%%%%%%%%%%%%%%%%%%%%%%%%%%%%%%%%%%%%%%%%%%%%%%%%%%%%%%%%%%%%%%
\section{Plasma properties in magnetic neutron stars}
%%%%%%%%%%%%%%%%%%%%%%%%%%%%%%%%%%%%%%%%%%%%%%%%%%%%%%%%%%%%%%%%%%%%%%%%

The transport properties and in general all physical properties 
of dense matter are strongly affected by the presence
of intense magnetic fields. Before discussing in details the microscopical
properties of magnetic matter, we begin by reminding some typical definitions
and quantities that serve as indicators of the relative importance of the
magnetic field.

The pressure in the crust and envelope is dominated by the contribution of the 
degenerate electrons.  Consider an electron gas whose number density is $n_e$. In the
absence of magnetic field, the Fermi momentum $p_{F}$, or equivalently the wave number 
$k_{F} = p_{F}/\hbar$ is
\beq
k_{F} = ( 3\pi^{2} n_{e} )^{1/3} = \left( \frac{3\pi^{2} \rho Z}{A m_{u}} \right)^{1/3}
\eeq 
where $m_u$ is the atomic mass unit, and we have assumed that the ions,
with atomic number $Z$ and atomic weight $A$, are completely ionized. This assumption
allows to relate $n_{e}$ and the density $\rho$, $\rho=\frac{n_e}{Z}A m_u$.
Defining the dimensionless quantity:
\beq
x_F = \frac{\hbar k_{F}}{m_e c} = 0.010066 \left( \frac{\rho Z}{A} \right)^{1/3}~,
\eeq
the Fermi energy is $\epsilon_{F} = m_{e} c^{2} \sqrt{1+x_{F}^{2}} $ and the Fermi temperature is
$T_{F} = (\epsilon_{F}-m_{e}c^{2})/k_{B} = m_{e}c^{2} (\sqrt{1+x_{F}^{2}}-1)$. 
If the matter is at temperature $T$, the electrons are degenerate when $T \ll T_{F}$. This
condition is fulfilled in the whole NS except for the outermost parts.

The magnetic field affects the properties of all plasma components, specially the electron component. 
Motion of free electrons perpendicular to the magnetic field is quantized in Landau levels,
which produces that
the thermal and electrical conductivities (as well as other quantities) exhibit quantum oscillations.
These oscillations change the properties of the degenerate electron gas in the limit of strongly 
quantizing field in which almost all electrons populate the lowest Landau level. 
The electron cyclotron frequency corresponding to a magnetic field $B$ is given by
\beq
\omega_B = \frac{eB}{m_e c},
\eeq
and the magnetic field will be considered {\it strongly quantizing} if the temperature of the
electrons is $T \ll T_{B}$ and the density $\rho < \rho_{B}$, where
\beq
T_{B} &=& \frac{\hbar \omega_{B}}{k_{B}} \approx 1.34 \times 10^{8} \frac{B_{12}}{\sqrt{1+x_{F}^{2}}} ~\mbox{K}
\label{temb}
\\
\rho_{B} &=& \frac{A m_{u} n_{B}}{Z} \approx 7.045 \times 10^{3} \left(\frac{A}{Z}\right) 
B_{12}^{3/2}~\mbox{g/cm$^{3}$}
\label{rhob}
\eeq
Here $k_B$ is the Boltzmann constant and $n_{B} = (eB/\hbar c)^{3/2}/ (\pi^{2} \sqrt{2})$
is the electron number density at which the Fermi energy reaches the lowest Landau level. 
The magnetic field is called {\it weakly quantizing} if $T \leq T_{B}$ but $\rho \gtrsim \rho_{B}$.
In this case the quantum oscillations are not very
pronounced and occur around their classical value.
The oscillations disappear for $T \gg T_{B}$ or $\rho \gg \rho_{B}$ and the field can be
treated as classical.

Let us turn now to the properties of the ions.
In the absence of magnetic field, the physical state of the ions depends on the Coulomb parameter
\beq
\Gamma = \frac{(Z e)^{2}}{k_{B} T a_{i}} \approx
\frac{0.23~Z^{2}}{T_{6}} \left( \frac{\rho}{A} \right)^{1/3}
\label{gamma}
\eeq
where $a_{i} = (3/4 \pi n_{i})^{(1/3)}$ is the ion-sphere radius, and $\rho_6$ and $T_6$ are,
respectively, the density and temperature in units of $10^6$ \mbox{g/cm$^{3}$} and $10^6$ K.
When $\Gamma < 1$ the ions form a Boltzmann gas, when $1 \leq \Gamma < 175$ their state is
a coupled Coulomb liquid, and when $\Gamma \geq 175$ the liquid freezes into a Coulomb lattice.
In general, the quantization of the ionic motion will be significant for temperatures
lower than the Debye temperature, which is approximately (for ions arranged in a bcc lattice)
\beq
T_D \approx 0.45 \frac{\hbar \omega_{p_i}}{k_{B}}
\approx 3.5 \times 10^{3} \left( \frac{Z}{A} \right) \rho^{1/2}~\mbox{K}
\eeq
and $\omega_{p_i}$ is the ion plasma frequency
\beq
\omega_{p_i} = \left( \frac{4 \pi Z^{2} e^{2} n_{i}}{m_{i}} \right)^{1/2}.
\eeq

In presence of strong magnetic fields, the electrons in an atom are confined to the lowest 
Landau level, the atoms are elongated and with larger binding energy and covalent bonding
between them. Therefore, below some critical temperature (depending on the composition and
the magnetic field strength), the gaseous layers of highly magnetized
neutron stars may undergo a phase transition that turns the gas into liquid or solid state
depending on the value of the Coulomb parameter $\Gamma$ \citep{Lai01}.
For typical magnetic field strengths of $10^{13}$ G, a Fe atmosphere will
condensate for $T< 0.1$ keV while a H atmosphere needs temperatures lower than
0.03 keV to undergo the phase transition to a condensed state.
In such a condensed neutron star surface made of nuclei with atomic number $Z$
and atomic weight $A$, the pressure vanishes at a finite density
\begin{equation}
\rho_s \approx 560~A Z^{-3/5} B_{12}^{6/5} {\rm g\, cm}^{-3}
\label{rhos}
\end{equation}
where $B_{12}$ is the magnetic field in units of $10^{12}$ G.

In this latter case, matter is in solid state and phonons become an important agent to transport 
energy.  When $T \geq T_D$, many thermal phonons are excited in the lattice, and the phonons
behave as a classical gas. However, if temperature is low, $T < T_D$, phonons behave
as a Bose quantum gas and the number of thermal phonons is strongly reduced.
Therefore, the Debye temperature $T_D$ allows to discriminate the quantum behaviour from the classical one.
Another important parameter related to the phonon processes of scattering is the so-called Umklapp
temperature, $T_U=T_D Z^{1/3} e^2/(3\hbar v_F)$, with $v_F$ being the Fermi velocity of electrons.

\begin{figure}
\resizebox{\hsize}{!}{\includegraphics{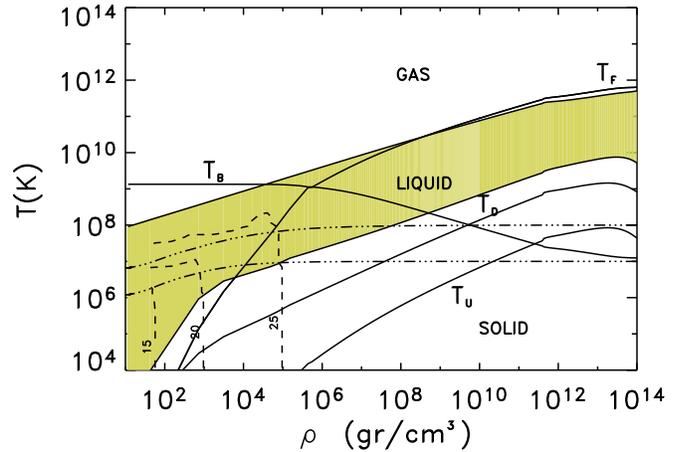}}
\caption{Phase diagram of neutron star matter.
The solid lines show the characteristic temperatures $T_{F}$, $T_{B}$, $T_{D}$, and $T_{U}$,
where $T_{F}$, $T_{B}$ have been calculated
for a magnetic field strength of $B = 10^{13}$G. 
The dashed lines show the domains of partial ionization (at $B=0$). 
The dot-dashed lines are two realistic temperature profiles for two different models with 
core temperatures of $10^{8}$ K and $10^{7}$ K.}
\label{fig1}
\end{figure}

All the previous properties and definitions are visualized and quantified in Fig. \ref{fig1},
where we show, in a phase diagram for neutron star matter,
the Fermi temperature ($T_F$) and $T_B$ (at $B = 10^{13}$ G), the Debye temperature ($T_D$) and
the Umklapp temperature. The central shaded band indicates
the region where matter is in the liquid state, according to Eq. (\ref{gamma}) and the condition
$1\le \Gamma< 175$. The region below that is the solid state ($\Gamma\ge 175$) and the region 
above that corresponds to the gaseous state ($\Gamma\le 1$). 
For reference, we have included two realistic 
temperature profiles (dot-dashed lines), corresponding to two different core temperatures
($10^{7}$ \mbox{K} and $10^{8}$ \mbox{K}).
The dashed lines appearing on the left--lower corner indicate the transitions to different
ionization states (25, 20 and 15 free electrons per atom). The outer layers are composed
of pure iron ($Z=26$). For $B = 10^{13}$ G, Eq. (\ref{rhob}) gives $\rho_B=5\times 10^5$
g/cm$^{3}$, therefore, the field is strongly quantizing only at low densities, weakly
quantizing in most of the envelope and classical in the crust ($T_B \le T$).
The zero pressure density as defined by Eq. (\ref{rhos}) is $7\times 10^4$ g/cm$^{3}$.

%%%%%%%%%%%%%%%%%%%%%%%%% FIGURE %%%%%%%%%%%%%%%%%%%%%%%%%%%%%%%%%

%%%%%%%%%%%%%%%%%%%%%%%%%%%%%%%%%%%%%%%%%%%%%%%%%%%%%%%%%%%%%%%%%%%%%%%
\section{Magnetic field structure.}

Although there is robust observational evidence that the external magnetic field is well 
represented by a dipolar configuration, the internal structure of the magnetic field 
in neutron stars is unknown, so that one has the freedom to prescribe arbitrary configurations. 
For weak magnetic fields (magnetic force negligible relative to the pressure gradient or
gravity) the deformation of the star is very small and the particular field structure 
is not important. 
Finding consistent numerical solutions of the Einstein-Maxwell equations describing the
structure of neutron stars endowed with a strong magnetic field, including
the effects of the Lorentz force and the curvature of the spacetime induced by
the stress-energy tensor of the magnetic field, is a difficult problem only
solved for purely poloidal configurations \citep{BBGN95} or very
recently including toroidal magnetic fields as perturbations \citep{IS04}. 
In previous works it has been shown that to obtain a significant deformation
of the star magnetic fields of the order of $10^{16}$ G are required. In this
work, for simplicity, and partially justified by the fact that most of our
models will be force-free (in a Newtonian sense) and less strong 
($10^{13}-10^{14}$ G), we will consider a spherical neutron star
described by a spherically symmetric metric of the form
\beq
ds^{2} = - e^{2 \Phi(r)} dt^{2} + e^{2 \Lambda(r)} dr^{2} +r^{2} (d\theta^{2}+\sin^{2} \theta d\phi^{2})
\eeq
where $e^{\Phi(r)}$ is the lapse function, $e^{\Lambda} = (1-2m/r)^{-1/2}$ 
is the space curvature factor, and we have taken $G=c=1$. The usual equation 
of hydrostatic equilibrium of a relativistic self-gravitating fluid is
\beq
\frac{dP}{dr} = - (\rho+P) \frac{d\Phi}{dr}  ~,
\label{hydeq}
\eeq
where $P$ is the pressure and $\rho$ is the mass-energy density.

Previous works about the effects of internal magnetic fields restrict the calculations 
to the simplest models (homogeneous, core dipole, purely radial in a thin layer), mainly
to simplify the problem. 
We address to the review by Tsuruta (1998), (section 5) for an overview
of the effects of the magnetic field on the thermal structure and evolution.

However, it is known that neutron stars are born
with differential rotation and that convective instabilities play a significant role
during the early stages of evolution \citep{KJM96,MPU00,MPU02}. Both differential rotation and 
convective motions should lead to non-trivial magnetic field structures with non-zero 
toroidal components \citep{BRU03}, that will evolve according to
\beq
\frac{\partial \vec{B}}{\partial t} = - \vec{\nabla} \times 
\left( \frac{c^{2}}{4 \pi} \hat{R} \cdot \vec{\nabla} \times 
\vec{B} \right)
\label{meeq}
\eeq 
where we have used the Newtonian equations for simplicity.
The relativistic versions of the induction equation for non-rotating and
rotating neutron stars can be found in the literature \citep{GPZ00,RA04}.
Above, $\hat{R}$ is the resistivity tensor.

In the classical (non quantizing) relaxation time approximation, 
the conductivities are related between them 
through the magnetization parameter ($\omega_{B} \tau_0$) where 
$\tau_0$ is the non-magnetic relaxation time \citep{UY80}.
Then, the magnetic field evolution equation (eq. \ref{meeq}) can 
be written as follows
\beq
\frac{\partial \vec{B}}{\partial t} = - 
\vec{\nabla} \times \left( \frac{c^2}{4 \pi \sigma_{\parallel}} \left( \vec{\nabla} \times \vec{B} 
+ \frac{\omega_{B} \tau_{0}}{B} \left( \vec{\nabla} \times \vec{B} \right) \times \vec{B} \right) \right)
\label{evolb}
\eeq
with $\sigma_{\parallel}$ being the electrical 
conductivities parallel to the magnetic field.
The first term at the right hand side of the above equation describes Ohmic dissipation 
and the last term is the Hall-drift, which is not dissipative but affects the current configurations.
In general, for strong magnetic fields, $\omega_{B} \tau_{0} \gg 1$, the Hall-drift cannot be neglected. 
Notice that 
\beq
\frac{c^2 \omega_{B} \tau_{0}}{4 \pi B \sigma_{\parallel}} = 
\frac{c}{4 \pi e n_e} 
\eeq
which does not depend on the relaxation time, making evident the non-dissipative character
of the Hall term. Even if the initial magnetic field is purely poloidal, 
it will develop a toroidal part during the evolution \citep{NK94,HR04,RKG04,CAZ04} and it is 
necessary to consider how it affects the thermal structure properties of neutron stars.
In the remaining of this section we describe the magnetic field configurations used in this work.

%%%%%%%%%%%%%%%%%%%%%%%%%%%%%%%%%%%%%%%%%%%%%%%%%%%%%%%%%%%%%%%%%
\subsection{Dipolar magnetic fields}
In some previous works, the structure of the magnetic field has been assumed to be poloidal,
in which case the field can be conveniently described in terms of
the Stokes stream function \citep{GU94,Mi98,Page00,GKP04}.  In spherical coordinates, and writing the 
$\phi$ component of the vector potential as $A_{\phi} = S(r) \sin \theta /r$,
the magnetic field components can be written in terms of the Stoke's function $S(r)$ as follows
\beq
B_{r} &=& \frac{2 S(r,t)}{r^{2}} \cos{\theta}  \nonumber \\
B_{\theta} &=& -\frac{\sin {\theta}}{r} \frac{\partial S(r,t)}{\partial r} ~.
\eeq

If we consider the static solution obtained by extending the vacuum solution 
to the center of the star we have
\beq
B_{r} = B_{0} R^{3} \frac{ \cos {\theta}}{r^{3}} \nonumber \\
B_{\theta} = \frac{B_{0} R^{3}}{2} \frac{ \sin {\theta}}{r^{3}} 
\eeq
which corresponds to the choice $S(r)=B_0 R/r$. This solution diverges at $r=0$.
For a magnetic field confined to the crust, $S(r)$ should vanish in the 
core due to proton superconductivity. 
In this work we prefer not to restrict ourselves to
use poloidal configurations and we use a more general structure
as described in the next subsection.

%%%%%%%%%%%%%%%%%%%%%%%%%%%%%%%%%%%%%%%%%%%%%%%%%%%%%%%%%%%%%%%%%
\subsection{Force-free magnetic fields}
A different, less restrictive, way to prescribe the interior magnetic field is to consider
a family of force-free fields. A force-free field is the simplest model for the equilibrium magnetic
field in the solar corona, above an active region of sunspots \citep{ff1,ff2}.
This class of fields are normally used to model the pre-flare coronal configurations
and have the advantage to allow large fields and currents to exist simultaneously
without exerting any force on the material. This helps to simplify the problem
because it allows to use the spherical solution
for hydrostatic equilibrium without magnetic fields and because
this configurations are not subject to the Hall drift.

A force-free magnetic field obeys
\beq
\vec{\nabla} \times \vec{B} = \mu \vec{B}
\label{maxwell2}
\\
\vec{B} \cdot \vec{\nabla} \mu = 0
\label{maxwell3}
\eeq
where the second equation is a result of the first one and the Maxwell's equation
$\vec{\nabla} \cdot \vec{B} = 0.$

Let us consider an axially symmetric magnetic configuration. In spherical coordinates,
with $(\theta,\phi)$ being the angular coordinates with respect to the magnetic field
axis of symmetry, a general magnetic field can be written in terms of the $\phi$ components
of the potential vector and the magnetic field as follows:
\beq
\vec{B} = \left( \frac{1}{r \sin \theta} \frac{\partial(\sin \theta A_{\phi})}{\partial \theta},
-\frac{1}{r} \frac{\partial(r A_{\phi})}{\partial r}, B_{\phi}\right)
\eeq
Therefore, Eq. (\ref{maxwell2}) reads:
\beq
\left( \frac{\sin \theta}{r} \frac{\partial(\sin \theta B_{\phi})}{\partial \theta},
-\frac{1}{r} \frac{\partial(r B_{\phi})}{\partial r}, 
\frac{1}{r} \left( \frac{\partial(r B_{\theta})}{\partial r}-\frac{\partial B_{r}}{\partial \theta} \right)
\right)
\nonumber \\
= \mu \left( \frac{\sin \theta}{r} \frac{\partial(\sin \theta A_{\phi})}{\partial \theta},
-\frac{1}{r} \frac{\partial(r A_{\phi})}{\partial r}, B_{\phi}\right).
\label{eqff}
\eeq
For simplicity, we will consider solutions with $\mu=$constant, so that 
Eq. (\ref{maxwell3}) is automatically satisfied, although more general solutions exist.
The equality of the $r,\theta$ components in Eq. (\ref{eqff}) is obviously satisfied
if we take $B_\phi = \mu A_\phi$. 
By analogy with the core dipole, we try the ansatz $A_\phi=\sin\theta ~ A(r)$, which leads
to the following equation for the $\phi$ components of Eq. (\ref{eqff})
\beq
\frac{d^{2} A(r)}{d r^{2}} + \frac{2}{r} \frac{d A(r)}{d r} 
 + \left(\mu^2 -  \frac{2}{r^{2}} \right) A(r) = 0 ~.
\eeq
This is a form of the Riccati-Bessel equation for $l=1$, which has solutions of
the form
\beq
A(r) = a j_{l}(x) + b n_{l}(x)
\label{bescom}
\eeq
where $a$,$b$ are constants, $x=\mu r$ and $j_{l}(x)$ and $n_{l}(x)$ 
are spherical Bessel functions of the first and second kinds. 
For $l=1$ we have, explicitly:
\beq
j_1(x) &=& \frac{\sin{x}}{x^2} - \frac{\cos{x}}{x}~,
\nonumber \\
n_1(x) &=& -\frac{\cos{x}}{x^2} - \frac{\sin{x}}{x}~.
\eeq
The spherical Bessel functions of the first kind are regular
in the origin ($j_l(x)\propto x^l$), while the functions of the second
kind diverge as $n_l(x)\propto x^{-(l+1)}$.

Hence, a general interior solution that matches 
(continuity of the normal component of the magnetic field)
with the vacuum dipolar solution at the surface is
\beq
\vec{B} = C \left( 2 \frac{\cos \theta}{r} A(r), -\frac{\sin \theta}{r} 
          \frac{\partial( r A(r))}{\partial r},
          \mu {\sin \theta} A(r) \right)
\label{bform}
\eeq
where $C=\frac{R_{S} B_0}{2 A(R_{S})}$ and $B_0$ is the value of the magnetic
field at the pole.
This magnetic field can be obtained from the following potential vector
\beq
\vec{A} = C \left( \mu r~{\cos \theta}~A(r), ~0,
          {\sin \theta}~A(r) \right)
\eeq

\begin{figure}
\resizebox{\hsize}{!}{\includegraphics{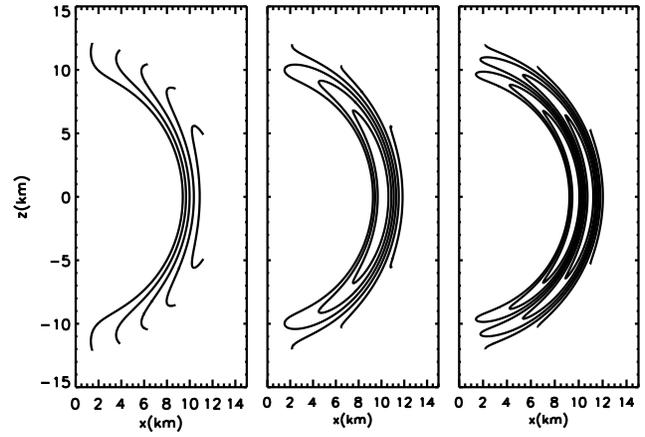}}
\caption{Projections of the field lines on the $(r,\theta)$ plane for force-free
magnetic field configurations confined in the region between $R_{int}=9.2$ km and 
$R_S=12.247$ km. From left to right $\mu=0.577, 1.569, 2.591$ km$^{-1}$. 
All configurations match continuously to an external dipole.}
\label{bessel}
\end{figure}

Notice that this general solution includes all simple configurations as particular limits.
The core dipole can be recovered by taking the limit $\mu \rightarrow 0$ and considering
only the $n_1(x)$ functions ($a=0$ in Eq. \ref{bescom}). Alternatively, in the limit
$\mu \rightarrow 0$ but considering only the family of regular solutions $j_1(x)$, we
arrive to the homogeneous magnetic field. In both cases, the $B_\phi$ component vanishes.
We can also find solutions that match continuously with the two components of the exterior
dipole by setting, $l=1$, $a = \cos (\mu R_{S})$, and $b=\sin (\mu R_{S})$. 
The family of solutions is parametrized
by the value of $\mu$, which can be interpreted as a wavenumber. If we want to build crustal
magnetic fields that match with an external dipole we only need to adjust the wavenumber
to have a vanishing radial component in the crust-core interface ($r=R_{\rm int}$) and 
continuity of $B_r$ and $B_\theta$ at the surface ($r=R_{S}$).
These values of $\mu$ are the solutions of
\beq
\tan \left(\mu \left(R_{\rm int} - R_{S} \right) \right) = \mu R_{\rm int}~.
\label{root}
\eeq
In Fig. \ref{bessel} we show three examples of crustal magnetic fields for the first 
three solutions of Eq. (\ref{root}) with $R_{\rm int}=9.2$ km and $R_{S}=12.247$ km.

In principle there are more solutions of Eq. (\ref{eqff}) than the linear one
($B_\phi = \mu A_\phi$) that we have adopted. A more general discussion about force-free
configurations can be found in the literature \citep{ff1,ff2}.

Finally, we also want to mention that this general force-free solution can be readily extended 
to higher order multipoles (i.e. quadrupole). This can be done by replacing in Eq. (\ref{bform})
$\cos\theta$ and $\sin\theta$ by the corresponding Legendre polynomial and its
derivative, and using the spherical Bessel functions of the same index $l$.
Another advantage of this force-free solution is that, for constant electrical resistivity
(although in a realistic NS this approximation is not appropriate), one can readily estimate
the diffusion time.
The evolution equation (\ref{evolb}) is simplified to
\beq
\frac{\partial \vec{B}}{\partial t} 
= - \frac{1}{\tau_{\rm diff}} \vec{B} 
\eeq
which solution reads
\beq
\vec{B} = \vec{B}_0~e^{-t/\tau_{\rm diff}}~.
\eeq
Here, $\vec{B}_0$ is the initial magnetic field and the decay time 
is $\tau_{\rm diff}=\frac{4\pi \sigma_{\parallel}}{c^2 \mu^2}$, that for typical
conditions in young neutron stars is $10^6-10^7$ years.

%%%%%%%%%%%%%%%%%%%%%%%%% FIGURE %%%%%%%%%%%%%%%%%%%%%%%%%%%%%%%%%

%%%%%%%%%%%%%%%%%%%%%%%%%%%%%%%%%%%%%%%%%%%%%%%%%%%%%%%%%%%%%%%%%%%%%%%%
\section{Thermal diffusion in highly magnetized neutron stars}
%%%%%%%%%%%%%%%%%%%%%%%%%%%%%%%%%%%%%%%%%%%%%%%%%%%%%%%%%%%%%%%%%%%%%%%%

\subsection{Equations.}
For very slowly rotating NSs, and neglecting the magnetic
force (which vanishes for force-free fields) the thermal evolution of 
neutron stars can still be described by the energy balance equation
\beq
C_{v} e^{\Phi (r)} \frac{\partial T}{\partial t} + \vec{\nabla} \cdot (e^{2 \Phi (r)} \vec{F}) = 
e^{2 \Phi (r)}\dot{\epsilon}
\label{eneq}
\eeq
where $C_{v}$ is the specific heat (per unit volume), $\vec{F}$ is the energy flux
and the source term ($\dot{\epsilon}$) includes all energy losses and sources
(neutrino emission, frictional or accretion heating, etc).
The evolution equation can also be written in integral form applying Gauss' theorem
\beq
\int_{V} e^{\Phi (r)} C_{v} \frac{\partial T}{\partial t} dV + \oint_{S} e^{2 \Phi (r)} \vec{F} \cdot d\vec{S} 
=\int_{V} e^{2 \Phi (r)} \dot{\epsilon} dV
\eeq

In the diffusion limit, the energy flux is given by 
\beq
e^{\Phi (r)}\vec{F} = - \hat{\kappa} \cdot \vec{\nabla} (e^{\Phi (r)} T)
\label{fluxeq}
\eeq
where $\hat{\kappa}$ is the thermal conductivity tensor. 
Defining a new variable $\tilde{T} = e^{\Phi (r)} T$, the 
components of the flux can be written as follows
\beq
e^{\Phi (r)} F_{r} = - (\kappa_{rr} e^{-\Lambda} \partial_{r} \tilde{T} + 
                         \frac{\kappa_{r \theta}}{r} \partial_{\theta} \tilde{T}) \nonumber \\
e^{\Phi (r)} F_{\theta} = - (\kappa_{\theta r} e^{-\Lambda} \partial_{r} \tilde{T} + 
                         \frac{\kappa_{\theta \theta}}{r} \partial_{\theta} \tilde{T})
\eeq
where $\kappa_{ij}$ are the components of the thermal conductivity tensor.
The $\phi$ component of the flux is not considered because we assume axial symmetry.

In the presence of strong magnetic fields, the thermal conductivities are different
in the directions parallel and perpendicular to the magnetic field. In the classical relaxation
time approximation, and considering that
only electrons carry heat, the ratio between the parallel and perpendicular conductivities
is related to the magnetization parameter ($\omega_B \tau_{0}$) 
as follows \citep{UY80}
\beq
\frac{\kappa_{\parallel}}{\kappa_{\perp}} = 1 + (\omega_{B} \tau_{0} )^{2}~.
\label{anis}
\eeq
The heat conductivity tensor in spherical coordinates and with the polar axis
coinciding with the axis of symmetry of the magnetic field can be written as follows
\begin{eqnarray}
&\hat{\kappa}& = \kappa_{\perp} \times
\nonumber \\
&&\left( I+(\omega_{B} \tau_{0})^2 \left( \begin{array}{ccc} 
 b_{r}^{2}  & b_{r} b_{\theta} & b_{r} b_{\phi} \\
 b_{r} b_{\theta}  & b_{\theta}^{2} & b_{\theta}b_{\phi} \\
 b_{r} b_{\phi}  & b_{\theta} b_{\phi} & b_{\phi}^{2} \\
               \end{array} \right) +  %\right. \nonumber \\
  \omega_{B}\tau_{0} \left(  \begin{array}{ccc}
      0      & b_{\phi} & -b_{\theta} \\
  -b_{\phi}  &    0     &  b_{r}      \\
   b_{\theta}& -b_{r}   &   0    \end{array} 
\right) \right) 
\nonumber \\ 
\end{eqnarray}
where $I$ is the identity matrix and $b_{r}, b_{\theta}, b_{\phi}$ are 
the components of the unit vector in the direction of the magnetic field.

With the above expression for $\hat{\kappa}$, the flux reads
\beq
e^{\Phi(r)} \vec{F} = - \kappa_{\perp} 
\left[ \vec{\nabla} \tilde{T} + \left(\omega_{B} \tau_{0} \right)^{2} \left( \vec{b} 
\cdot \vec{\nabla} \tilde{T} \right) 
\cdot \vec{b} + \omega_{B} \tau_{0} \left( \vec{b} \times \vec{\nabla} \tilde{T} \right) \right]
\label{flux}
\eeq
The Hall contribution to the heat flux is given by the last term on 
the right hand side of Eq. (\ref{flux}). If the magnetic field geometry 
has only  poloidal components, and the temperature distribution does not depend on the 
azimuthal angle, $\phi$, the divergence of the Hall term vanishes \citep{GKP04} and
it does not affect the energy balance equation (\ref{eneq}).
However, for a magnetic field structure with a toroidal component, this term 
contributes to the heat flux, even in axial symmetry.
In the following, to simplify notation, we will omit the {\it tilde} over the 
temperature and we will use the symbol $T$ for the {\it red-shifted} temperature.

%%%%%%%%%%%%%%%%%%%%%%%%%%%%%%%%%%%%%%%%%%%%%%%%%%%%%%%%%%%%%%%%%%%%%%%
\subsection{Boundary conditions.}

Boundary conditions can be imposed in either the temperature or the flux in the
boundaries of our numerical domain. Only a few years after birth the inner core
of a NS becomes isothermal, therefore, in the core-crust interface ($r=R_{int}$) 
we will impose a fixed core temperature ($T_c$).
At the surface we impose 
\beq
F(B,T,\theta_{B}) = \alpha(B,T,\theta_{B}) \sigma T^{4}
\label{ffactor}
\eeq
where $\sigma$ is the the Stephan-Boltzmann constant and $\alpha(B,T,\theta_{B})$ is 
the integrated emissivity that depends on the particular model,
and $\theta_{B}$ is the angle between the magnetic field and the 
direction normal to the surface element.
At the temperatures of interest and for magnetic fields intense enough to produce
the condensation of the gaseous layers, the emissivity at low energies is strongly
reduced compared to the blackbody case 
and depends on the orientation angle \citep{Bri80,TZD04,paper1,Lai05}. 

\begin{figure}
\resizebox{\hsize}{!}{\includegraphics{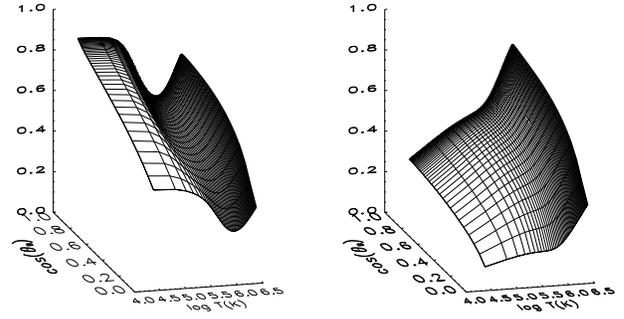}}
\caption{Flux factor $\alpha$ as defined in Eq. (\ref{ffactor}) for a dipolar magnetic field
with $B_{p} = 5 \times 10^{13} G$, with (left panel) and without (right panel) taking into account 
the effect of the motion of the ions (considered as free particles).}
\label{ajust1}
\end{figure}

In Fig. \ref{ajust1} we show the flux factor with (left) and without (right) taking into account the 
effect of the motion of ions for a dipolar magnetic field of $B_0 = 5 \times 10^{13} G$.
Based on our previous results \citep{paper1},
we have obtained a polynomial fit of $\alpha(B,T,\theta_{B})$ as a function of $T_6$
(temperature in units of $10^6$ K) and 
$\cos (\theta_{B})$ for different magnetic field strengths (relative error $< 2 \%$)
with the following form:
\beq
\alpha = \sum_{i=1}^{6} \sum_{j=1}^{6} a_{i,j} T_6^{i-1} \cos^{j-1}(\theta_{B})
\label{alfafit}
\eeq
The $a_{i,j}$ coefficients for $B=10^{13}$ G and $5\times10^{13}$ G with and without
taking into account the effect of the motion of ions are presented in 
tables (\ref{ion513tab}--\ref{wion13tab}).
The case of isotropic emission (black body) can be recovered by setting $\alpha=1$.

%%%%%%%%%%%%%%%%%%%%%%%%% FIGURE %%%%%%%%%%%%%%%%%%%%%%%%%%%%%%%%%

%%%%%%%%%%%%%%%%%%%%%%%%%%%%%%%%%%%%%%%%%%%%%%
\subsection{Numerical test.}
\begin{figure}
\resizebox{\hsize}{!}{\includegraphics{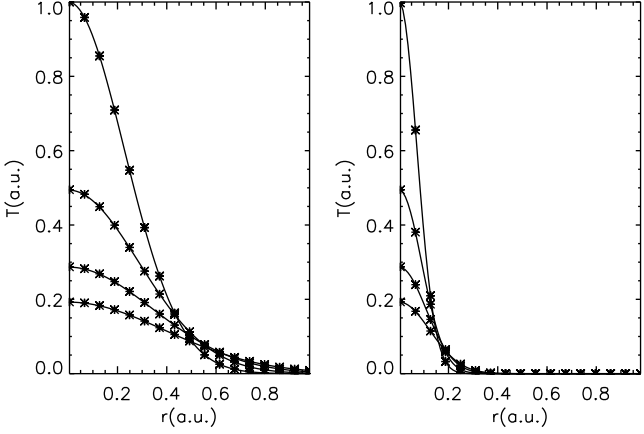}}
\caption{Temperature profiles at different times comparing the analytic solution
(solid) and the numerical evolution (stars) of a thermal pulse in a medium
embedded in a homogeneous magnetic field. For clarity, in the numerical solution
we have shown only one out of every six grid points. The parameters of the simulation
are $\omega_B \tau_0=3$ and $\kappa_{\perp} = 10^2$ (a.u), $t_{0}=10^{-4}$.
The left panel shows the evolution of polar profiles and the right panel
corresponds to equatorial profiles.}
\label{aevol}
\end{figure}

The numerical algorithm consists of a standard finite difference scheme fully implicit in time.
The temperature is cell-centered, while the fluxes are calculated at each cell-edge.
In order to test the code, we have studied the evolution of a thermal pulse 
in an infinite medium (neglecting all general relativistic effects), embedded in a
homogeneous magnetic field oriented along the z-axis. 
If the conductivity is constant in the medium,
an analytical solution for
the temperature profile at a time $t$ is the following:
\beq
T(\vec{r}, t) = T_0 {\left( \frac{t_0}{t} \right)}^{3/2} 
\exp{\left[ -\frac{r^{2}}{4 \kappa_{\perp} t} \left(\sin^{2} \theta + \frac{\cos^{2} \theta}
{1+ (\omega_B \tau_0)^2}
\right) \right]}
\label{ansol}
\eeq
where $T_{0}$ is a constant (the central temperature at the initial time, $t_0$), 
$\kappa_{\perp}$ is the transverse conductivity 
and $(\omega_B \tau_0)$ is the magnetization parameter.
To check the accuracy of the method, we have compared the numerical evolution of the pulse with the 
analytical solution, for different values of the parameters $\kappa_{\perp}$ and $\omega_B \tau_0$. 
As boundary conditions, we prescribe the temperature corresponding to the analytical solution
in the surface and we impose $\vec{F}=0$ at the center.
In Fig. \ref{aevol} we show the comparison between the analytical (solid) and numerical
(stars) solution and for a model with $\omega_B \tau_0=3$ and $\kappa_{\perp} = 10^2$ (a.u).
The grid resolution is 100$\times$40 (radial$\times$angular).
The deviations from the analytical solution in all cases studied
are less than $0.1 \%$.

%%%%%%%%%%%%%%%%%%%%%%%%%%%%%%%%%%%%%%%%%%%%%%%%%%%%%%%%%%%%%%%%%%%%%%%

%%%%%%%%%%%%%%%%%%%%%%%%%%%%%%%%%%%%%%%%%%%%%%%%%%%%%%%%%%%%%%%%%%%%%%%
\section{Microphysics input.}

The microphysical ingredients that enter in the transport equations (\ref{eneq}) and
(\ref{fluxeq}) are the specific heat and the thermal conductivity.
Strictly speaking, the specific heat is not needed to obtain stationary configurations, 
but we have chosen to evolve Eq. (\ref{eneq}) without sources
with a fixed inner temperature until the stationary solution is reached. Therefore, 
by using realistic microphysics input we will obtain also information about the 
thermal relaxation timescales.

The dominant contribution to the specific heat is that from electrons and ions.
For electrons we use the formulae corresponding to a relativistic degenerate Fermi gas 
while for ions we follow van Riper (1991).  The most important ingredient is, however, 
the thermal conductivity, which has contribution from electron,
photon and phonon transport. In this section we summarize the expressions used in the simulations.

%%%%%%%%%%%%%%%%%%%%%%%%%%%%%%%%%%%%%%%%%%%%%%%%%%%%%%%%%%%%%%%%%%%%%%%
\subsection{Thermal conductivity.}

The region of interest covers a large range of densities,
from the core-crust boundary ($\approx 10^{14}$ g/cm$^3$) to the surface, which is given by 
Eq. (\ref{rhos}) in the models of condensed atmosphere. 
The total conductivity includes the contributions of three carriers,
\beq
\kappa = \kappa_e + \kappa_{rad} + \kappa_{ph}
\eeq
where $\kappa_{e}$ is the electron conductivity, $\kappa_{rad}$ is the radiative (photon) conductivity
and $\kappa_{ph}$ is the phonon conductivity.
In non magnetic neutron stars, heat is transported mainly by electrons in the crust and 
in the inner envelope and by photons near the surface, while the phonon transport is negligible. 
However, in the presence of strong magnetic fields, 
this situation may change. While for the transport along the magnetic field the phonon contribution is 
still negligible, in the transverse direction the electron transport is drastically suppressed, 
and the phonon contribution may become the most important one.
Let us consider each of this contributions separately.

%%%%%%%%%%%%%%%%%%%%%%%%%%%%%%%%%%% 
\subsubsection{Electron transport}
In the crust and the envelope of a neutron star the transport properties
are mainly determined by the process of electron scattering off strongly correlated
ions. The study of the transport properties of Coulomb plasmas with and without magnetic fields has
been a focus of attention for decades \citep[e.g.][]{FI76,UY80,KY81, Itoh84}.
For the envelope, we will use the expressions obtained by
Potekhin (1999), who calculated the thermal and electrical conductivities of degenerate electrons
in magnetized envelopes by means of an effective scattering potential that takes into account
multiphonon processes in Coulomb crystals and an appropriate structure factor of ions in Coulomb liquids. 
For the crust, the practical expression derived by Potekhin (1999) have been later generalized
\citep{GYP01} by including how the size and shape of nuclear charge affects the
transport properties as well as reconsidering the electron-phonon scattering processes.

In our calculations we are using the results from Potekhin (1999), whose code is
of public domain \footnote{{\tt www.ioffe.rssi.ru/astro/conduct/condmag.html}}.
For pedagogical purposes, in order to make evident the effect of a large magnetization
parameter, we now summarize the classical relations for degenerated electrons.
Schematically, the thermal conductivity can be written in terms of some
effective relaxation time, $\tau_{ij}$, as follows
\beq
%\sigma_{ij} &=&  \frac{e^{2} n_{e} c^{2}}{\epsilon_{F}} \tau_{ij}(\epsilon_{F}) \\
\kappa_{ij} &=&  \frac{\pi^{2} k_{B}^{2} n_{e} c^{2} T}{3 \epsilon_{F}} \tau_{ij}(\epsilon_{F}),
\eeq
where $\tau_{ij}$ are interpreted as inverse effective collision frequencies.  In the non-quantizing 
case, we can write explicitly the different components in terms of the magnetization 
parameter ($\omega_B \tau_0$)
\begin{eqnarray}
\tau_{zz} = \tau_0, \, \tau_{xx} = \frac{\tau_0}{1+\omega_{B} \tau_0}, \, 
\tau_{yx} = \frac{\omega_{B}\tau_0^{2}}{1+\omega_{B} \tau_0}
\end{eqnarray}

The three main electron scattering processes that play a role in our scenario are scattering
off ions, electron-phonon scattering and scattering off impurities.
Semi-analytic expressions and fitting formulae for the relaxation times and thermal
conductivities along the magnetic field for all three processes were derived 
by Potekhin \& Yakovlev (1996).
The total contribution of electrons to the thermal conductivity is then calculated as 
\beq
\kappa_e = \left( \kappa_{e-ph}^{-1}+\kappa_{e-imp}^{-1} \right)^{-1}~.
\eeq

%%%%%%%%%%%%%%%%%%%%%%%%% FIGURE %%%%%%%%%%%%%%%%%%%%%%%%%%%%%%%%%
\begin{figure}
\resizebox{\hsize}{!}{\includegraphics{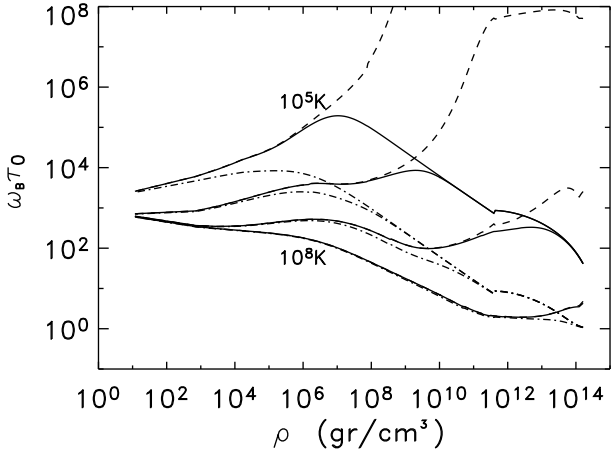}}
\caption{Magnetization parameter ($\omega_{B} \tau_{0}$) against density for
different temperatures (from top to bottom $10^5$,$10^6$,$10^7$,$10^8$ K) and $B=10^{13}$G. 
The solid lines are calculated with an impurity parameter of $Q=0.1$, the dashed lines are 
for homogeneous matter ($Q=0$) and the dot-dashed lines correspond to highly inhomogeneous 
matter $Q=10$.}
\label{emagn}
\end{figure}
%%%%%%%%%%%%%%%%%%%%%%%%%%%%%%%%%%%%%%%%%%%%%%%%%%%%%%%%%%%%%%%%%%

In Fig. \ref{emagn} we show the magnetization parameter, related to the anisotropy
of the thermal conductivity by Eq. (\ref{anis}), as a function of density and
for different temperatures.  For comparison, we show results with different impurity
concentration parameter $Q = {n_{\rm imp}(Z_{\rm imp}-Z)^{2}}/{n_{i}}$. 
The dashed, solid, and dash-dotted lines correspond to $Q=0, 0.1$, and 10,
respectively. For highly inhomogeneous matter ($Q=10$), the magnetization parameter
is strongly reduced in the crust ($\omega_B \tau_0 \approx 1$).

At large temperature the total electron conductivity is weakly dependent on temperature.
If the temperature drops below the Umklapp temperature ($T \ll T_{U}$), the Umklapp
processes are disallowed and $\kappa \propto T^{-4}$. Therefore, at high temperature
the dominant process is the electron-phonon scattering but at low temperature
the scattering off impurities becomes the dominant contribution.

%%%%%%%%%%%%%%%%%%%%%%%%%%%%%%%%%%%%%%%%%%%%%%%%%%%%%%%%%%5
\subsubsection{Photon transport}

Radiative conduction becomes the most effective transport mechanism in the 
outermost layers of the envelope, where electrons are non degenerate. 
We employ the expressions derived by Potekhin \& Yakovlev (2001) for
fully ionized iron, who fitted previous results \citep{Sil80}.

Free-free transitions and Thompson scattering off free electrons are the two 
contributions to the total radiative conductivity, that is calculated according to
\beq
\kappa_{rad} = \left( \kappa^{-1}_{ff}+\kappa^{-1}_T \right)^{-1}
\eeq
For temperatures below $10^{7}$ K, the dominant contribution comes from free-free transitions, 
which scales as $\approx \rho^{-2} T^{6.5}$. Notice that
for $T< 10^{7}$ K we have $\kappa_{\perp} \approx 2\kappa_{\parallel}$.

%%%%%%%%%%%%%%%%%%%%%%%%%%%%%%%%%%%%%%%%%%%%%%%%%%%%%%%%%%
\subsubsection{Phonon transport}

Energy transport by phonons is usually orders of magnitude less effective
than the usual electron or radiative transport. However, in the situation that
we are studying, the large anisotropy induced by the magnetic field can suppress
electron thermal conduction in the perpendicular direction by factors of $10^3-10^6$.
Under this circumstances, transport by phonons
become important, since this processes will become the most effective way to transport 
energy in the perpendicular direction. For this reason, we need to include it
in our calculations.

%%%%%%%%%%%%%%%%%%%%%%%%% FIGURE %%%%%%%%%%%%%%%%%%%%%%%%%%%%%%%%%
\begin{figure}
\resizebox{\hsize}{!}{\includegraphics{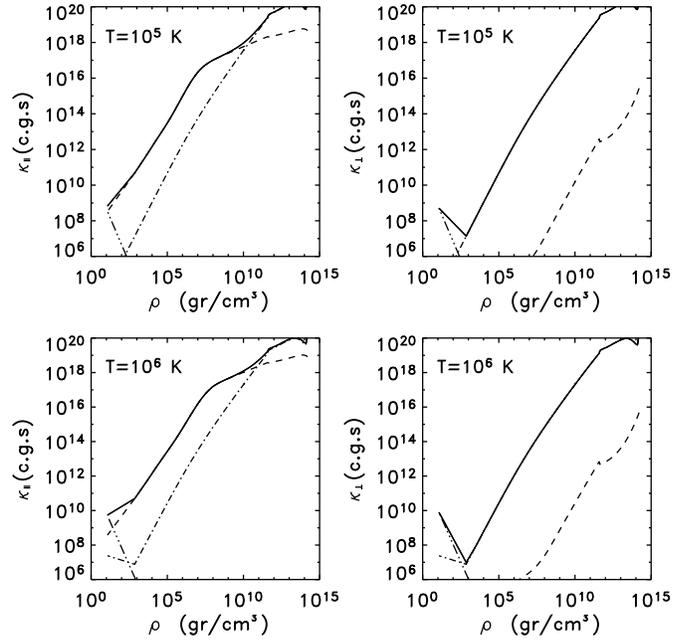}}
\caption{Thermal conductivity due to electron transport (dashed line), 
phonon transport (dot dashed line), 
photon transport (double dot dashed) and total (continuous line). 
The left panels show the conductivity in the direction longitudinal to the magnetic
field while the right panels show the transverse conductivity, which is strongly 
suppressed.}
\label{condtotal}
\end{figure}
%%%%%%%%%%%%%%%%%%%%%%%%%%%%%%%%%%%%%%%%%%%%%%%%%%%%%%%%%%%%%%%%%%

In a first approximation, we consider a very simplified model, in which the phonon distribution is 
characterized by a Debye spectrum and all the relaxation times are functions of the wave vector of 
one mode only. In this approximation,
the lattice thermal conductivity can be expressed as
\beq
\kappa_{ph} = 
\frac{k_{B}}{2 \pi^{2} c_{s}} \left( \frac{k_{B}}{\hbar} \right)^{3} T^{3} \int_{0}^{T_{D}/T} \tau(x)
\frac{x^{4} \exp{(x)}}{(\exp{(x)}-1)^{2}} dx
\label{peq1}
\eeq
where $c_{s}$ is the sound speed, $T_{D}$ is the Debye temperature, $x$ 
is a dimensionless variable ($x \equiv \hbar \omega / k_{B} T $) and $\tau$ is the combined relaxation time, 
whose reciprocal is the sum of the reciprocal relaxation times for all scattering processes 
considered, Umklapp and impurity scattering processes (both dissipative) and 
the three phonon normal scattering which are non dissipative \citep{Holl63,Konst}:
\beq
\tau^{-1} = \tau_{U}^{-1}+\tau_{I}^{-1}+\tau_{N}^{-1} 
\eeq

At temperatures $T>T_{D}$, the lattice conductivity is mainly determined by 
the Umklapp processes, and the integral (\ref{peq1}) can be approximated to the expression  
\beq
\kappa_{ph} \sim \frac{\rho c_{s}^{3} a}{4 T}
\label{peq2}
\eeq
where $a \approx \left( \frac{A m_{B}}{\rho} \right)^{1/3}$ is the lattice constant. 
At lower temperatures ($T<T_{D}$), dissipative processes make the conductivity to 
increase very rapidly and the inclusion of impurity and normal phonon scattering becomes 
necessary. These processes (which conserve the total momentum) cannot by 
themselves lead to a finite thermal conductivity, but do not allow very large heat currents 
to be carried by modes of long wavelength. In this low temperature limit the thermal 
conductivity can be expressed in the form \citep{Call61}
\beq
\kappa_{ph} = \frac{k_{B}}{2 \pi^{2} c_{s}} \left( \frac{k_{B}}{\hbar} \right)^{3} T^{3}
\frac{\left( \int_{0}^{T_{D}/T} x^{4} \exp{(x)}(\exp{(x)}-1)^{-2} dx \right)^{2}}
{\int_{0}^{T_{D}/T} \tau_{D}^{-1} x^{4} \exp{(x)}(\exp{(x)}-1)^{-2} dx}
\label{peq3}
\eeq
where $\tau_D^{-1} = \tau_{U}^{-1}+\tau_{I}^{-1}$. In the limit $\tau_{U}^{-1}=0$
we recover the Ziman limit \citep{Ziman}
\beq
\kappa_{ph} = \frac{2 \hbar \omega_D^3}{360 \pi^{3} c_{s} T \Gamma_{\rm imp}}
\eeq
where $\omega_D$ is the Debye frequency 
and $\Gamma_{\rm imp}$ is a parameter that takes into account the different atomic
masses of the impurities and the lattice deformation.

According to this expressions, in the crust and inner envelope, the 
heat transport along the magnetic field is dominated by the electron component, 
while phonons become the main transport agent in the transverse direction
(see Fig. \ref{condtotal}). If the core temperature is low enough ($T<10^5$K), the 
phonon contribution becomes very important also in the parallel direction, 
making the crust to be quasi isothermal.
Near the surface, the radiative conductivity dominates in both directions. 

%%%%%%%%%%%%%%%%%%%%%%%%% FIGURE %%%%%%%%%%%%%%%%%%%%%%%%%%%%%%%%%
\begin{figure*}
\resizebox{\hsize}{!}{\includegraphics{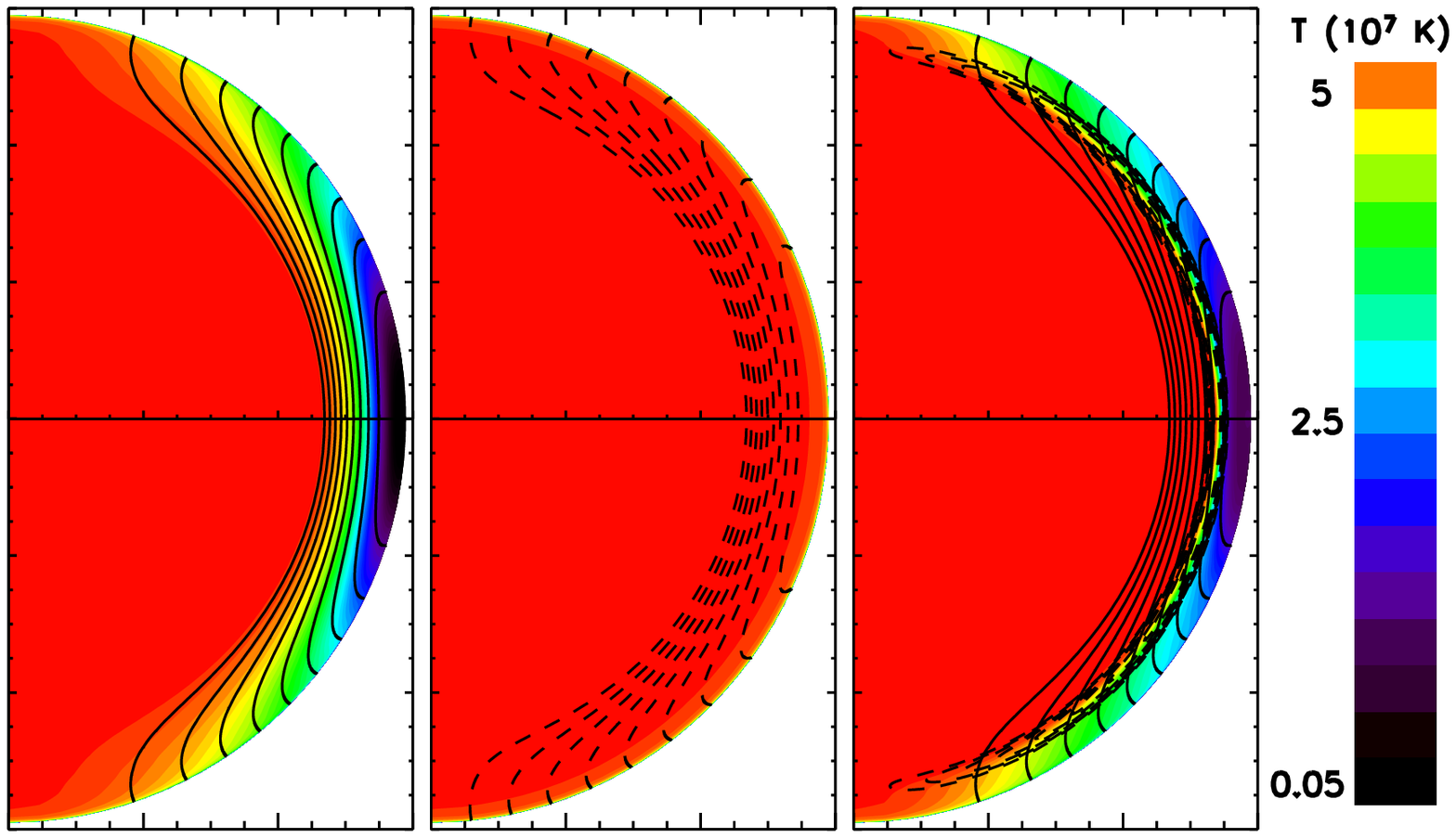}}
\resizebox{\hsize}{!}{\includegraphics{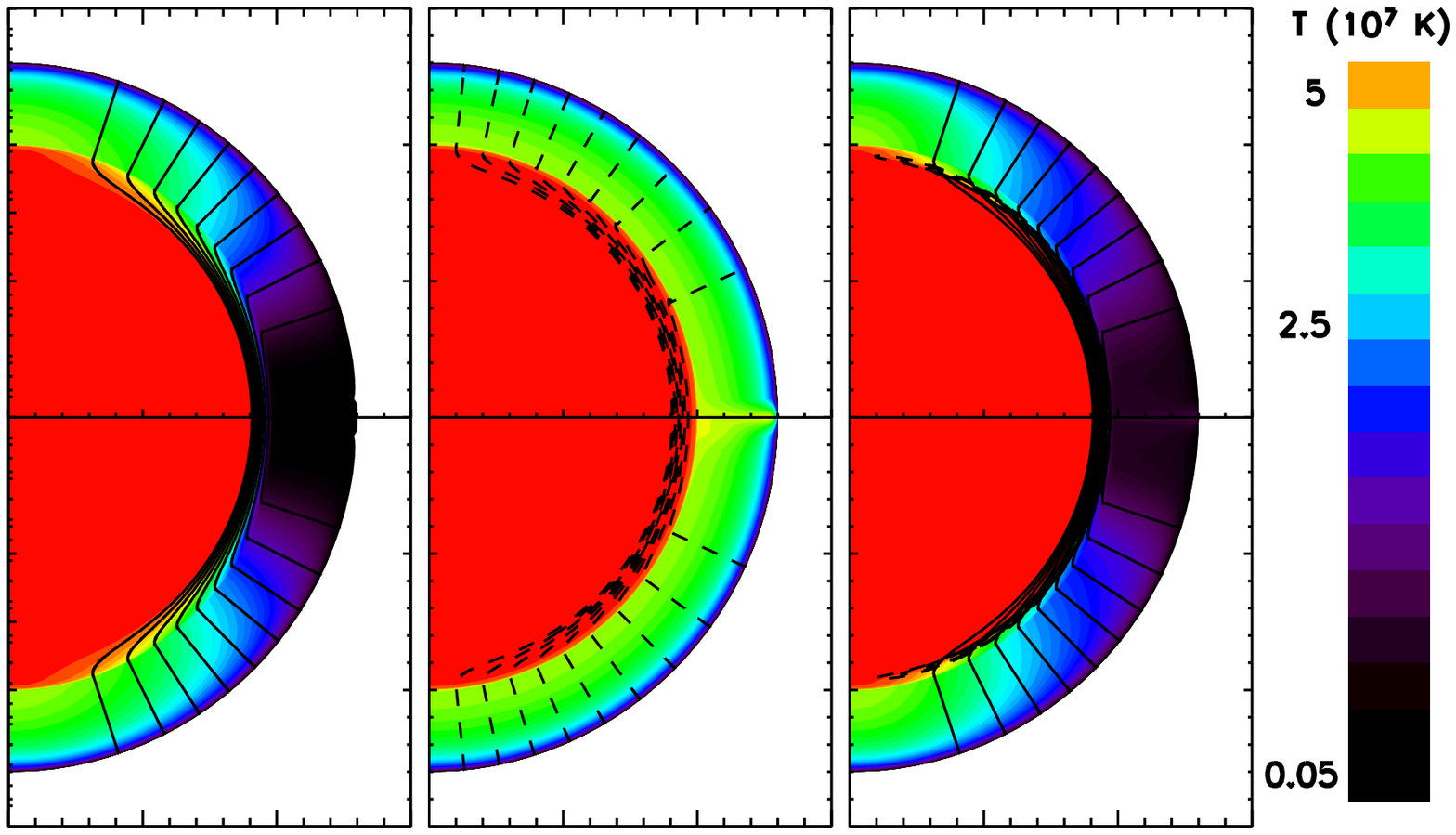}}
\caption{Upper panel: Temperature distribution in the crust of neutron stars with different
toroidal components. The poloidal component is the same in all models
($B_p=10^{13}$ G) and it is confined to the crust
(solid lines). The left panel shows results for a purely poloidal field, the central
panel a force-free configuration, and the right panel corresponds to
a toroidal component confined to a narrow region of the crust.
In the two latter cases the dashed lines show contours of constant
$B_{\phi}$. The scale has been stretched about a factor 2 to enlarge the crustal region.
Lower panel: Same as the upper panel but stretching the scale to enlarge 
the region of the envelope.
}
\label{fccrust}
\end{figure*}

%%%%%%%%%%%%%%%%%%%%%%%%%%%%%%%%%%%%%%%%%%%%%%%%%%%%%%%%%%%%%%%%%%%%%%%
\section{Results.}

Our aim is to find stationary solutions of the temperature distribution
in a given background magnetic field configuration. 
We assume that the inner core is isothermal, and that the diffusion time 
of the magnetic field ($\tau_{\rm diff}$)
is much longer than the relaxation time to reach thermal equilibrium so that the
magnetic field is kept fixed. 
We also assume that the sources or sinks of energy (if any) are effective only 
at longer timescales. 
This assumptions are justified because both the magnetic diffusion time (either
Ohmic or ambipolar) and the cooling time are $>10^5$ years, while the typical time
to achieve the stationary solution (starting from a constant temperature profile)
is $\approx 10^3$ years. Notice that the 
diffusion timescale when the Hall instability occurs is about $10^4$ years, 
so that in this case one would need to consider the coupled evolution of the 
temperature and the magnetic fields.
Instead of solving the equation $\vec{\nabla} \cdot \vec{F} = 0$ directly,
we evolve Eq. (\ref{eneq}), without sources, 
until the stationary solution is reached. 

The main effect of the magnetic field on the temperature distribution 
can be guessed by looking at the expression of the heat
flux (\ref{flux}). When the magnetization parameter is large ($\omega_B \tau_0 \gg 1$),
the dominant contribution to the flux is proportional 
to $(\omega_B \tau_0)^2 (\vec{B} \cdot \vec{\nabla} T)$. 
Therefore, in order to reach  the stationary configuration the temperature distribution
must be such that the surfaces of constant temperature are practically aligned with the 
magnetic field lines ($\vec{B} \cdot \vec{\nabla} T \ll 1$).
This is shown explicitly in the left panel of Fig. \ref{fccrust}, where we show the stationary 
solution for a purely poloidal configuration confined to the crust and the outer layers (
poloidal confined, PC in the following).
This alignment is enforced in most of the crust and envelope, and only near the surface 
strong radial gradients are generated. When we introduce a toroidal component the situation
changes, because the Hall term in Eq. (\ref{flux}) induces large meridional fluxes
(order $\omega_B \tau_0$) which result in an almost isothermal crust. 
This is clearly seen in the central panel, that shows the temperature
distribution for a force-free magnetic field (FF) with a toroidal component present in the
outer layers (crust and envelope). For comparison, we also considered another non force-free
model (right panel) which has a toroidal component
confined to a thin crustal region (toroidal confined, TC in the following), with a maximum
value of $2\times10^{15}$G.  It acts as an insulator keeping
a different temperature at both sides of the toroidal field. In the region external to
the toroidal field, since only the poloidal component is present we see again the alignment
of isothermal surfaces with the magnetic field lines, which would not happen if the $B_{\phi}$
component extend all the way up to the surface, as in the central panel.
We must stress again that the poloidal field is the same
in all three models (solid lines, $B_p=10^{13}$ G), but the field lines have been omitted 
in the central panel for clarity.
The dashed lines are contours of constant $B_{\phi}$. In the lower panel of Fig. \ref{fccrust} 
we show the same results but stretching artificially the low density regions to make visible
the gradients near the surface. A slight north-south asymmetry provoked by the Hall term
is visible in the right panel.

%%%%%%%%%%%%%%%%%%%%%%%%% FIGURE %%%%%%%%%%%%%%%%%%%%%%%%%%%%%%%%%
\begin{figure}
\resizebox{\hsize}{!}{\includegraphics{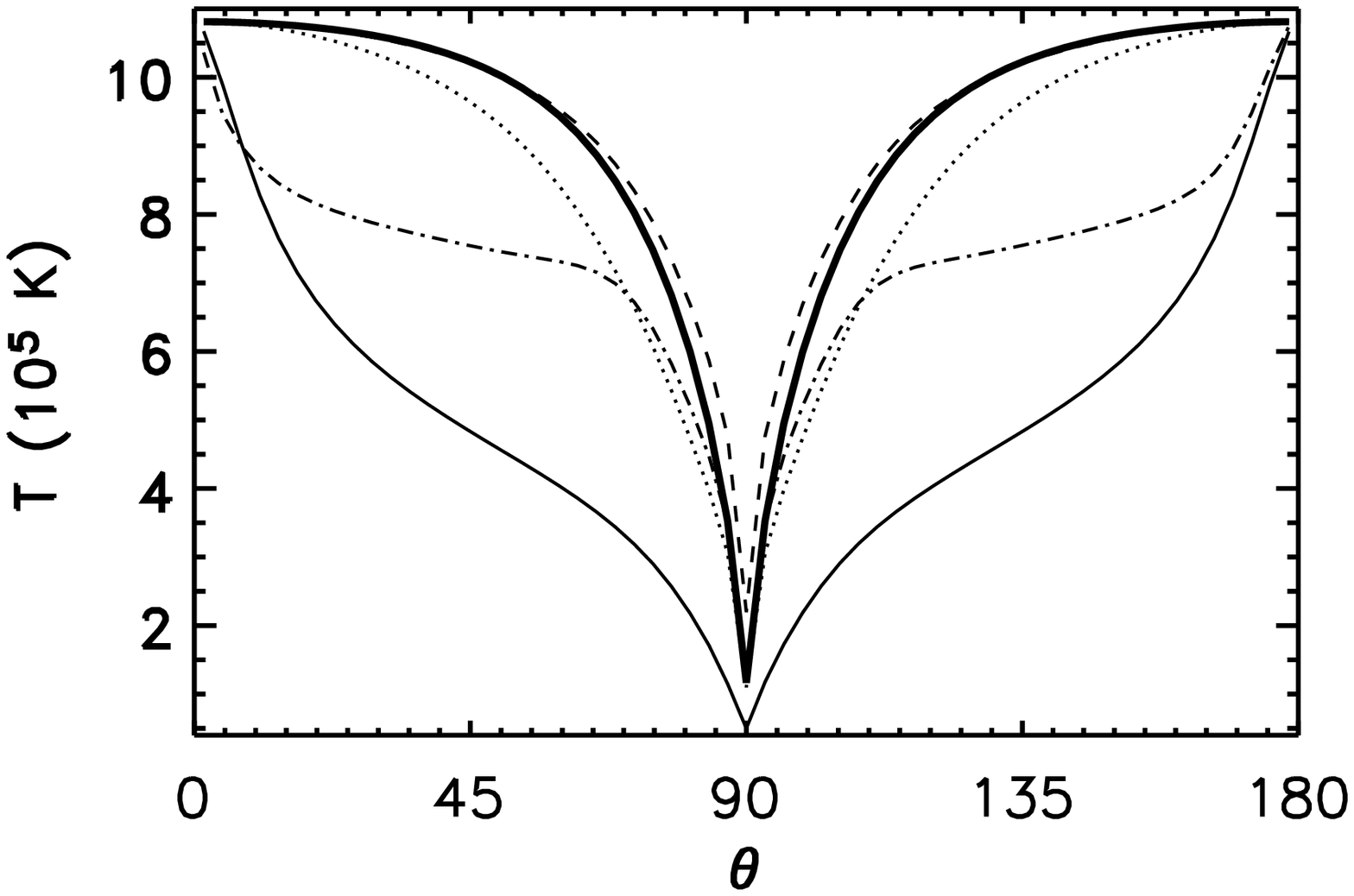}}
\caption{Surface temperature profiles as a function of the polar angle for different
magnetic field configurations with the same value at the pole  $B_p=10^{13}$ G.
The core temperature for all models is 5$\times 10 ^7$ K.
The models considered are: core dipolar (dashed), PC (dotted),
TC (dash-dot), and FF (thin solid line). In all cases we have
included phonon transport effects with $\Gamma_{\rm imp}=0.1$.
The temperature distribution of
Greenstein and Hartke (\ref{GHf}) is also shown for comparison (thick solid line).
}
\label{fsurf}
\end{figure}

%%%%%%%%%%%%%%%%%%%%%%%%% FIGURE %%%%%%%%%%%%%%%%%%%%%%%%%%%%%%%%%
\begin{figure}
\resizebox{\hsize}{!}{\includegraphics{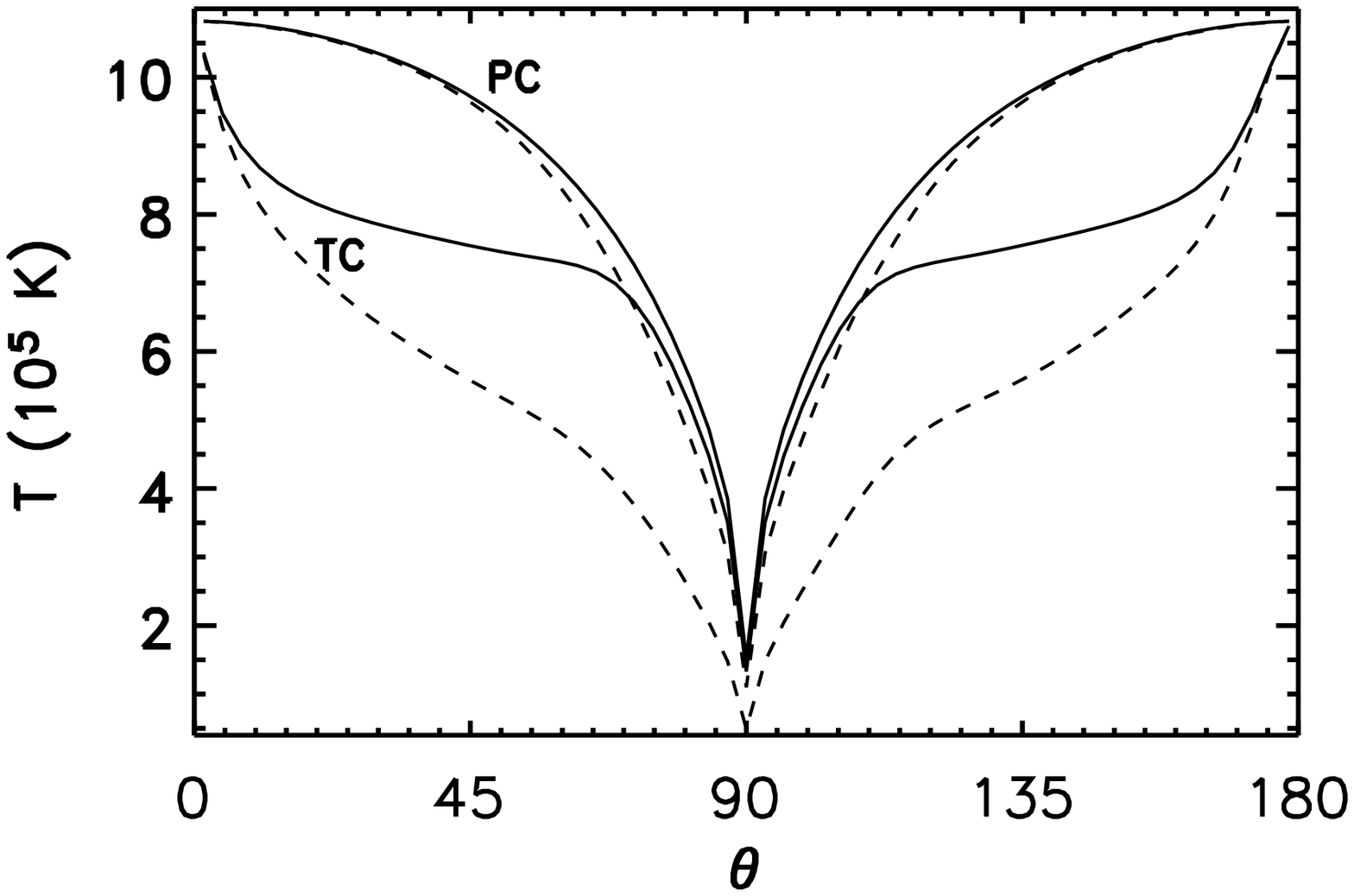}}
\caption{Surface temperature profiles as a function of the polar angle for 
the PC and TC configurations. The dashed lines have been calculated without
taking into account the phonon contribution, while the solid lines 
correspond to models including phonon transport effects with an 
impurity parameter of $\Gamma_{\rm imp}=0.1$.
}
\label{fsurfph}
\end{figure}
The core temperature for all models is 5$\times 10 ^7$ K.
Thus, the anisotropy induced by the field becomes important not only in the crust but
also in the condensed envelope. The direct consequence is a non-uniform surface 
temperature distribution shown in Fig. \ref{fsurf}, where we show the angular 
distribution of the surface temperature for several magnetic field configurations, 
all of them with the same surface magnetic field (dipolar, $B_p=10^{13}$ G).
For comparison, we have also included (thick solid line)
the semi-analytic temperature distribution derived by {Greenstein and Hartke} (1983),
\beq
T^{4} = T_{p}^{4} \left( \cos^{2}{\theta_{B}} 
+ \frac{\kappa_{\perp}}{\kappa_{\parallel}} \sin^{2}{\theta_{B}} \right)
\label{GHf}
\eeq
where $\theta_{B}$ is the angle between the normal vector to the surface and the magnetic field. 
The figure compares the following models: core dipole (dashed line), PC (dotted line), 
FF (thin solid line) and TC (dash-dot). Qualitatively, the purely poloidal configurations 
(core dipole, PC) look similar to 
the Greenstein \& Hartke solution of Eq. (\ref{GHf}), with quantitative
differences of the order of 10-20\%. The general structure (relatively large hot polar region, 
and narrow cool equatorial band) is reproduced by all models without toroidal components of the
magnetic field. This situation changes when a toroidal magnetic field is included, as for example
in the force free configuration. In models with important toroidal components, the surface thermal
distribution consists of a small hot polar region and a relatively large cooler (about a factor 2-3)
area. The slight north-south asymmetry provoked by the Hall term
is only visible in the TC model (compare thin solid line with dot-dashed line).

In the models where large meridional gradients in the crustal region are not present
the phonon contribution to the thermal  conductivity is not relevant and varying
the impurity concentration barely changes the results. This is not true for the most
extreme models (PC, TC), where phonon transport can make a difference.
In Fig. \ref{fsurfph} we compare the surface temperature distribution in the PC and
TC models when phonon transport is switched off.
The solid lines correspond to models in which the phonon contribution to
the thermal conductivity is included (we have taken $\Gamma_{\rm imp}=0.1$), 
while the dashed lines show models obtained without including the phonon conductivity. 
Despite the fact that the effect of phonons is evident in the TC model, we must point 
out that the general distribution (small hot polar cap, larger cooler region)
remains similar.

A simple way to understand the results presented in this section is based
on the following arguments. The models can be generally classified in two
subclasses: i) magnetic field configurations that result in almost isothermal
crusts (core dipole, FF, homogeneous) and ii) configurations for which large
crustal temperature gradients are present (PC, TC).
The first subclass includes models without toroidal components but also models
with toroidal components present in the whole crustal region (i.e. FF). 
As discussed at the beginning of this section, the Hall term in Eq. (\ref{flux})
is responsible of the meridional heat flux that smears out temperature
anisotropies in the crust. For such models, the surface temperature distribution
is well reproduced by the classical Greenstein and Hartke formula 
(\ref{GHf}) but noticing that the dependence of $\theta_B$ with the polar angle
$\theta$ is different for each model. For a core dipole, we have
\beq
\cos^{2}{\theta_B}=\frac{4 \cos^{2}{\theta}}{1+3\cos^{2}{\theta}}~,
\eeq
for a FF model
\beq
\cos^{2}{\theta_B}=\frac{4 \cos^{2}{\theta}}{(1+\mu^2 R^2)
+(3-\mu^2 R^2) \cos^{2}{\theta}}
\eeq
and for a homogeneous magnetic field $\cos^{2}{\theta_B}=\cos^{2}{\theta}$.
We have checked that $T=T(\theta_B)$ looks very similar in all three cases
despite the apparent differences in the surface distribution $T(\theta)$.
The size of the hot polar cap can be easily estimated for this models.
If we define the angular size of the polar cap as the angle where the temperature
has decreased a given factor (say a factor 2, for example) with respect to
the polar temperature, we can solve for $\theta$ to obtain the polar cap size.

Models that admit strong crustal temperature gradients (PC, TC) do not obey
Eq. (\ref{GHf}), and in principle there is no simple way to obtain how the
temperature  varies with the polar angle. The only general rule is that
a strong toroidal component is necessary to produce small hot polar caps.

%%%%%%%%%%%%%%%%%%%%%%%%%%%%%%%%%%%%%%%%%%%%%%%%%%%%%%%%%%%%%
\subsection{Effective temperature.}

\begin{figure}
\resizebox{\hsize}{!}{\includegraphics{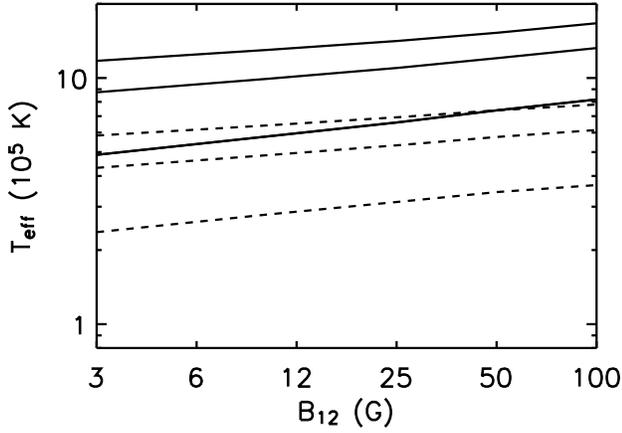}}
\caption{Dependence of the effective temperature on the core temperature
and the magnetic field strength for two different configurations
core dipole (solid) and force free (dashes).
We show results for three different core temperatures,
from bottom to top $10^7$, $5\times10^7$, and $10^8$ K.
}
\label{Teff}
\end{figure}

In Fig. \ref{Teff} we show the dependence of the effective temperature on the
core temperature and the magnetic field strength. The effective temperature is
defined as $L = 4 \pi R_S^2 \sigma T_{\rm eff}^4$, where $L$ is the
total integrated luminosity over the surface. This effective temperature is 
the quantity usually obtained from black-body fits to observational data, and
plotted on cooling curves to compare data with theoretical predictions.
The three solid lines correspond to three different core temperatures,
from bottom to top $10^7$, $5\times10^7$, and $10^8$ K, and for a core dipole
configuration. The dashed lines correspond to the same core temperatures but
for a force free magnetic field. In all cases we observe a systematic lower
effective temperature (a factor $\approx 2$) in configurations with toroidal
magnetic fields in the crust-envelope region. This means that among NS with similar
ages (i.e. similar core temperatures during the neutrino dominated cooling era),
those with strong toroidal fields have an apparent effective temperature about
a factor 2 smaller than those with low magnetic fields or purely dipolar configurations.

%%%%%%%%%%%%%%%%%%%%%%%%% FIGURE %%%%%%%%%%%%%%%%%%%%%%%%%%%%%%%%%

Our result can also be compared to the classical formula that relates
the temperature at the base of the envelope with the surface temperature \citep{GPE82}
\beq
T_{b,8} = 1.288 \left(\frac{T_{{\rm eff},6}}{g_{14}}\right)^{0.455} 
\eeq
where $T_{b,8}$ is the temperature in the base of the envelope in units of $10^8$ K,
$T_{{\rm eff},6}$ is the surface temperature in units of $10^6$ K and $g_{14}$ is the
gravity acceleration in units of $10^{14}$ c.g.s. For the three core temperatures
($10^7, 5\times 10^7, 10^8$ K)
used in Fig. \ref{Teff}, and assuming an isothermal crust, the surface temperature
of non-magnetized NSs would be of $0.05, 1.85$ and $8.58\times 10^5$ K, respectively.
The quite different effective temperatures predicted by different models are relevant
for the interpretation of the comparison of observational data with cooling curves.

%%%%%%%%%%%%%%%%%%%%%%%%%%%%%%%%%%%%%%%%%%%%%%%%%%%%%%%%%%%%%
\subsection{Quantizing magnetic field effects.}

\begin{figure}
\resizebox{\hsize}{!}{\includegraphics{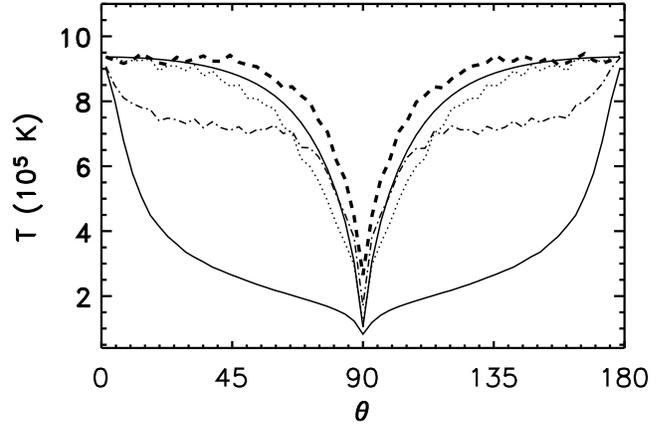}}
\caption{Surface temperature profiles as a function of the polar angle for 
the same models as in Fig. \ref{fsurf} but including quantizing effects
on the electron thermal conductivity.}
\label{fig9q}
\end{figure}

\begin{figure}
\resizebox{\hsize}{!}{\includegraphics{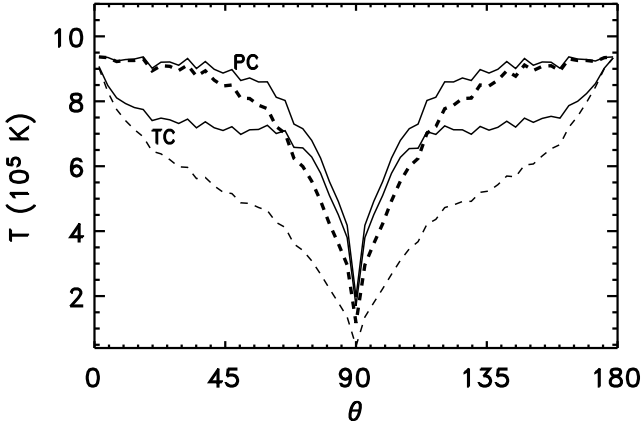}}
\caption{Same as Fig. \ref{fsurfph}
but including quantizing effects
on the electron thermal conductivity.}
\label{fig10q}
\end{figure}

\begin{figure}
\resizebox{\hsize}{!}{\includegraphics{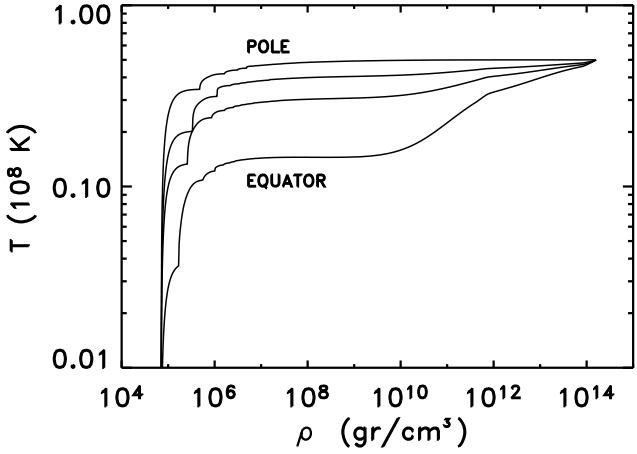}}
\caption{Temperature profiles as a function of the density for different
polar angles $\theta = 0, 45, 60$, and 90$^\circ$. The magnetic field configuration
is TC with $B_p=10^{13}$ G and a core temperature of 5$\times 10 ^7$ K.
}
\label{tprof}
\end{figure}

In the previous results the quantizing character of the magnetic field has
been neglected.
For $10^{13}$ G, Eq. (\ref{rhob}) gives
$\rho_B\approx 4.8 \times 10^5$ g/cm$^3$, while the density of the condensed
surface is $\rho_s \approx 7 \times 10^4$ g/cm$^3$. Therefore only in the outermost
thin layer ($\rho<\rho_B$) the magnetic field can be considered strongly quantizing 
while in most of the envelope it is weakly quantizing. 
Including the quantizing effects on the conductivities
in a transport code is challenging because the rich structure in Landau levels
makes necessary a robust code and high resolution to handle properly the
gradients that might develop near each transition. For a selected number of
models we have performed the calculations including quantizing effects
with the purpose of understanding the qualitative and quantitative differences with 
the classical case.
In Figs. \ref{fig9q} and \ref{fig10q} we show the results for the same models as 
in Figs. \ref{fsurf} and \ref{fsurfph}
but including quantizing effects on the electron thermal conductivity
(Potekhin, 1999).
The two main facts that we observe in this figures (present as well in other models 
not shown) are the following. First, the
average effective temperature is generally lower and the anisotropy is more pronounced,
i.e. a smaller angular size of the hot polar region.
Second, the surface temperature distribution shows small oscillations in models without toroidal 
components near the surface (core dipolar, PC, TC), associated to the oscillatory behaviour
of the thermal conductivity in the quantizing case.
This can be explained by the fact that the poloidal component is practically radial and heat 
transport in the meridional direction is strongly suppressed. 
The radial gradients are different at each latitude due to
the different magnetic field strength, and therefore different densities at which electrons 
are filling the corresponding Landau levels. This is shown in Fig. \ref{tprof} where
radial temperature profiles for three different polar angles are plotted. The Landau
levels are clearly visible. This oscillatory behaviour cannot be smoothed out by
meridional heat fluxes because they are suppressed by a factor $\approx (\omega_B \tau_0)^2$. 
The exception is the FF model, for which oscillations are not observed, because 
the it has a toroidal component extended up to the surface. This makes possible
the existence of heat flux in the meridional direction because the Hall term
is order $(\omega_B \tau_0) \kappa_\perp$.

%%%%%%%%%%%%%%%%%%%%%%%%% FIGURE %%%%%%%%%%%%%%%%%%%%%%%%%%%%%%%%%
%%%%%%%%%%%%%%%%%%%%%%%%% FIGURE %%%%%%%%%%%%%%%%%%%%%%%%%%%%%%%%%
%%%%%%%%%%%%%%%%%%%%%%%%%%%%%%%%%%%%%%%%%%%%%%%%%%%%%%%%%%%%%

%%%%%%%%%%%%%%%%%%%%%%%%% FIGURE %%%%%%%%%%%%%%%%%%%%%%%%%%%%%%%%%
%%%%%%%%%%%%%%%%%%%%%%%%% FIGURE %%%%%%%%%%%%%%%%%%%%%%%%%%%%%%%%%
\subsection{Influence of the physical conditions of the outer layers.}

\begin{figure}
\resizebox{\hsize}{!}{\includegraphics{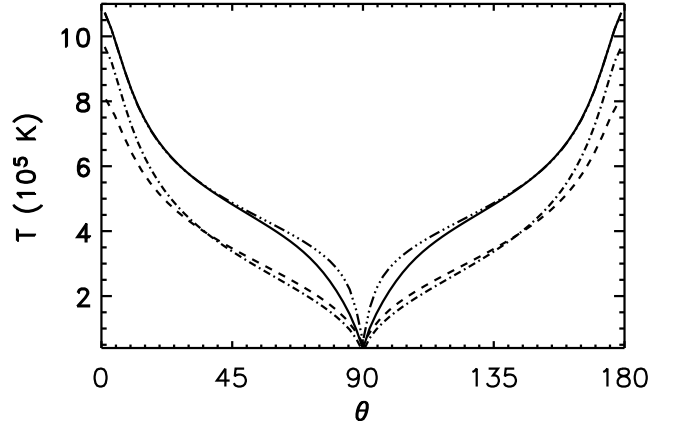}}
\caption{Influence of the boundary condition on the surface temperature distribution for 
a model with $T_{\rm c}=5\times10^{7}$K and a force-free magnetic field configuration with 
$B=10^{13}$G. Results are shown for blackbody emission
(dot-dashed), gaseous magnetic envelope (dashed), and metallic surface with (solid)
and without (triple dot - dash) taking into account the motion of the ions. 
}
\label{bcond}
\end{figure}

\begin{figure}
\resizebox{\hsize}{!}{\includegraphics{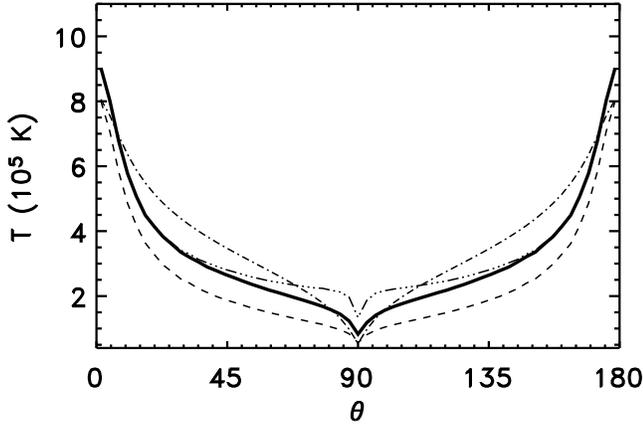}}
\caption{Same as Fig. \ref{bcond} but including quantizing effects
of the magnetic field.
}
\label{fig12q}
\end{figure}

One of the important issues under debate is whether the envelope and atmosphere
will be in a gaseous or condensed state. In order to estimate the dependence of our results
on the choice of a particular model, we have compared four different outer boundary conditions
which represent the different possibilities one can find. 
A common approach is to solve the 2D heat
transfer equation only in the crust \citep{GKP04} and match at 
$\rho \approx 10^{10}$ g/cm$^3$ to some magnetized 
envelope solution \citep{PY01}. Alternatively, as we have discussed in this work, 
there is the condensed surface model in which the emissivity at low energies may
vary depending on the way that the motion of ions in the lattice (fixed or free ions
are the two limits) is treated \citep{paper1,Lai05}. 
Also one can consider the simplest model which is to assume
that the condensed surface (e.g. $\rho_s=7\times 10^4$ g/cm$^3$, for $B=10^{13}$ G) 
radiates as a blackbody. 
We have analyzed this four possibilities and we show the resulting
surface temperature distribution for all four models in Figs. \ref{bcond} and \ref{fig12q},
for the classical and quantizing cases, respectively.
The main difference, as stated previously, is that including quantizing effects
leads to lower average temperatures. Notice also that when quantizing effects
are included, the size of the hot polar region is smaller and the temperature 
is nearly constant in a large part of the surface of the star. 

It must be stressed that these are all FF configurations, for which the crust is very close 
to isothermal, and the gradients of temperature are generated in the low density region.
The conclusion from this comparison is that not only the temperature at the base 
of the crust, but also the physical conditions in the low density layers affect the
total luminosity (the average effective temperature). However, the general shape of the 
surface temperature distribution is qualitatively similar in all cases, which
leads to conclude that irrespectively of the physical assumptions, neutron
stars with strong magnetic fields do have large surface temperature variations.

%%%%%%%%%%%%%%%%%%%%%%%%% FIGURE %%%%%%%%%%%%%%%%%%%%%%%%%%%%%%%%%
%%%%%%%%%%%%%%%%%%%%%%%%% FIGURE %%%%%%%%%%%%%%%%%%%%%%%%%%%%%%%%%

It a forthcoming work we will study models with stronger magnetic fields
(magnetar rather than isolated NS conditions) in which the quantizing effects are 
probably even more important.
For the remaining of this work, and having established the qualitative trends that 
differentiate classical and quantizing models, we will focus on analyzing, in the
classical limit, a number of other different issues that might have important consequences
on the emission properties.

%%%%%%%%%%%%%%%%%%%%%%%%%%%%%%%%%%%%%%%%%%%%%%%%%%%%%%%%%%%%%%%%%%%%%%%%%%%%%%%%%%5
\subsection{Influence of impurities.}

The influence that impurities or defects in the lattice have on the final temperature
distribution may be important. Impurity scattering dominates either at very low
temperatures (where phonon scattering is suppressed) or when the impurity level is very
high. For isolated NSs the values of the impurity parameter may
vary from $Q \approx 10^{-3}$ in very pure crusts to $Q \ge 10$ in the amorphous inner crust
\citep{Jon04}. In accreting neutron stars $Q$ is set by the composition of the nuclear
burning occurring at low density, and it is likely that $Q \approx 100$ \citep{Sch99}.
The impurity content also determines the critical field above which the Hall effect
dominates over purely Ohmic dissipation \citep{CAZ04}.
Given this uncertainty, we have explored a variety of models with the
impurity parameter to test the sensitivity of our results to the impurity concentration.

\begin{figure}
\resizebox{\hsize}{!}{\includegraphics{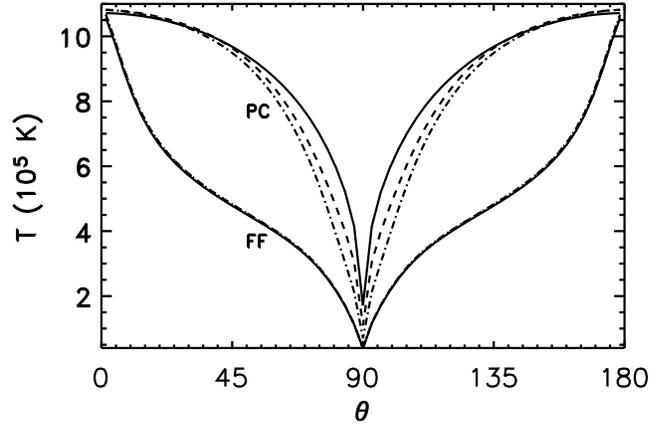}}
\caption{Influence of the impurity content for
models with $T_{\rm c}=5\times10^{7}$K and two different
magnetic field configurations (PC and FF), both with the same 
poloidal component ($B=10^{13}$ G). 
We show results for $Q=0$ (dash-dots), $Q=0.1$ (dashes) and $Q=10$ (solid).
}
\label{bimp}
\end{figure}

In Fig. \ref{bimp} we show the resulting surface temperature distributions corresponding to
the PC and FF configurations and for different values of
the impurity parameter $Q=0$ (dot-dashed), 0.1 (dashed), and 10 (solid).
Only small corrections to the PC configuration are visible, while the lines are 
indistinguishable for the FF model. Therefore, the exact value of the impurity 
concentration might be important for the long term evolution of the magnetic
field and the temperature, but it does not seem to be crucial for the 
stationary solution corresponding to a background field.

%%%%%%%%%%%%%%%%%%%%%%%%%%%%%%%%%%%%%%%%%%%%%%%%%%%%%%%%%%%%%%%%%%%%%%%%%%%
\begin{table*}
\caption{Maximum pulsed fraction for ${\cal O} = \pi/2$ and ${\cal B}=\pi/2$ with
a core dipolar configuration.}
\begin{tabular}{lccccc}
\hline
\hline
\noalign{\smallskip}
  & $B_{p}=3 \times 10^{12}$ G & $ B_{p}=10^{13}$ G & $B_{p}=2.5\times10^{13}$ G & $B_{p}=5\times10^{13}$ G& $ B_{p}=10^{14}$ G \\
\hline\noalign{\smallskip}
 $ T_{\rm c}=10^7$ K     &  0.15     & 0.14      & 0.13          & 0.11         & 0.10  \\
 $ T_{\rm c}=5\times10^7$ K &  0.16     & 0.16        & 0.16         & 0.15         & 0.13  \\
 $ T_{\rm c}=10^8$ K        &  0.16     & 0.16        & 0.16         & 0.16         & 0.15  \\

\hline\noalign{\smallskip}
\end{tabular}
\label{pfdip}
\end{table*}

\begin{table*}
\caption{Maximum pulsed fraction for ${\cal O} = \pi/2$ and ${\cal B}=\pi/2$ with
a force-free configuration.}
\begin{tabular}{lccccc}
\hline
\hline
\noalign{\smallskip}
  & $B_{p}=3 \times 10^{12}$ G & $ B_{p}=10^{13}$ G & $B_{p}=2.5\times10^{13}$ G & $B_{p}=5\times10^{13}$ G& $ B_{p}=10^{14}$ G \\
\hline\noalign{\smallskip}
 $ T_{\rm c}=10^7$ K     &  0.18     & 0.18        & 0.18     & 0.19      & 0.21  \\
 $ T_{\rm c}=5\times10^7$ K &  0.26  & 0.24        & 0.23     & 0.22      & 0.22  \\
 $ T_{\rm c}=10^8$ K     &  0.28     & 0.27        & 0.26     & 0.25      & 0.24  \\

\hline\noalign{\smallskip}
\end{tabular}
\label{pfff}
\end{table*}
%%%%%%%%%%%%%%%%%%%%%%%%%%%%%%%%%%%%%%%%%%%%%%%%%%%%%%%%%%%%%%%%%%%%

\subsection{Pulsations.}

\begin{figure}
\resizebox{\hsize}{!}{\includegraphics{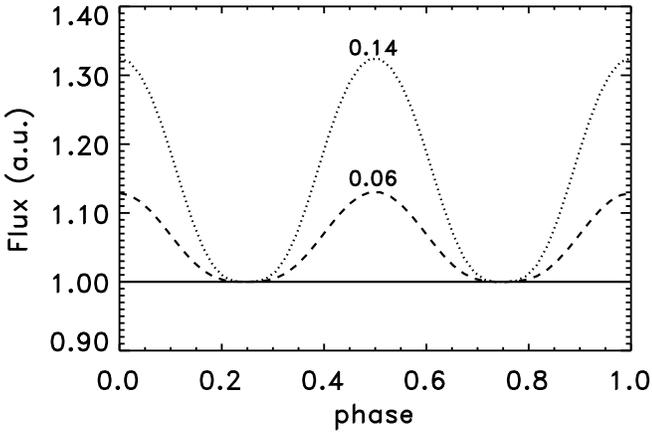}}
\caption{Observed flux variability and the corresponding maximum pulsed fraction
for three different orientations of the magnetic axis with
a fixed observer position (${\cal O} = \pi/2$):
solid (${\cal B} = 0$), dotted (${\cal B} = \pi/2$) and dashed
(${\cal B} = \pi/4$). The magnetic field structure is core dipolar with
$B_{p} = 10^{13}$ G and $T_{\rm c} = 10^{7}$ K.
}
\label{l9045}
\end{figure}

The anisotropic temperature distribution obtained from our calculations will translate
into periodic pulsations if the neutron star is rotating and the magnetic and rotation
axis are not aligned. Within a fully relativistic framework that includes light bending
effects \citep{Page95,Page96}, we have calculated the visible luminosity curves for
a number of models with different magnetic field strengths and configurations.
In Fig. \ref{l9045} we show the observed luminosity obtained for models
with a core dipolar configuration with $B_{p} = 10^{13}$ G and $T_{\rm c} = 10^{7}$ K.
We denote by ${\cal O}$ the angle between the observer and the rotation
axis and by ${\cal B}$ the angle between the rotation and magnetic axis.
The numbers next to each line are the maximum pulsed fraction (MPF):
\begin{equation}
{\rm MPF} = \left. \frac{F_{max}-F_{min}}
{F_{max}+F_{min}} \right|_{{\cal B} = \pi/2}
\end{equation}
The dependence of the MPF
on the different parameters can be understood by analyzing the results
summarized in Tables \ref{pfdip} (core dipolar configurations) and \ref{pfff}
(force free configurations). Several interesting conclusions can be drawn from
the results. First, for a given core temperature, it depends 
very weakly on the strength of the magnetic field ($B_p$), but can by up to twice
larger when the toroidal magnetic field is included (force free models).
For poloidal fields the MPF in the models we analyzed is 16\%,
but it can be increased to 25-30\% in the force-free models. We must remind that
the toroidal component is larger in about one order of magnitude than the
poloidal one (see Eq. (\ref{bform}), $\mu R_{S} \approx 10$). The anisotropy and therefore
the variability may be increased
by using other magnetic field configurations with larger toroidal components.
The observed variability of isolated neutron stars in consistent with this results,
but some of them show large pulse fractions (11\% in RX J0720, 12\% in RX J0420,
18\% in RBS 1223) than seem to indicate the existence of toroidal interior magnetic
fields, and large angles between the rotation and magnetic axis.
The lack of pulsations in (RX J1856) is compatible with nearly aligned rotation
and magnetic axis (${\cal B}<6^{\circ}$). The information about
the variability correlated with the effective temperature and the optical
excess flux can therefore give relevant information about the magnetic field structure.

%%%%%%%%%%%%%%%%%%%%%%%%%%%%%%%%%%%%%%%%%%%%%%%%%%%%%%%%%%%%%%%%%%%%%%%%%%%%%%%%

%%%%%%%%%%%%%%%%%%%%%%%%%%%%%%%%%%%%%%%%%%%%%%%%%%%%%%%%%
\subsection{A comparison to blackbody models.}

\begin{figure}
\resizebox{\hsize}{!}{\includegraphics{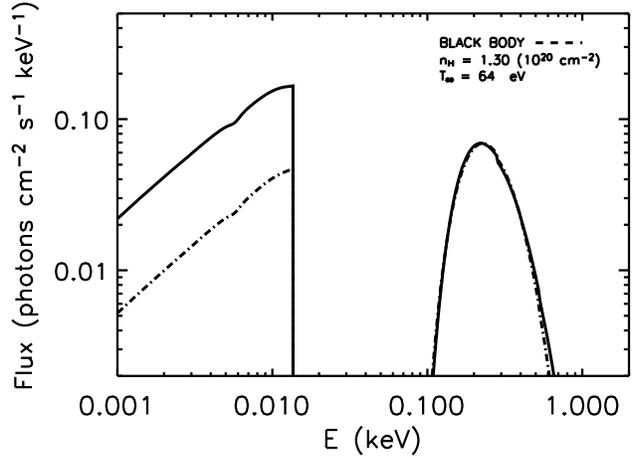}}
\caption{Comparison between the spectra of a FF model (solid line)
with $B_{p} = 10^{13}$G and $T_{\rm c} = 5 \times 10^{7}$ K and a single
blackbody fit (dashed line). The parameters of the {\it real} NS are
$M=1.4 M_\odot$, $R=12.27$ km, and we have taken $d=117 pc$
and $n_H=1.5\times 10^{20}$cm$^{-2}$.
We have assumed ${\cal O} = \pi/2$ and ${\cal B} = 0$.
The parameters of the BB fit are given in the figure.
The optical flux of the FF model is a factor 4.3 larger than the BB fit.
}
\label{fit1}
\end{figure}

\begin{figure}
\resizebox{\hsize}{!}{\includegraphics{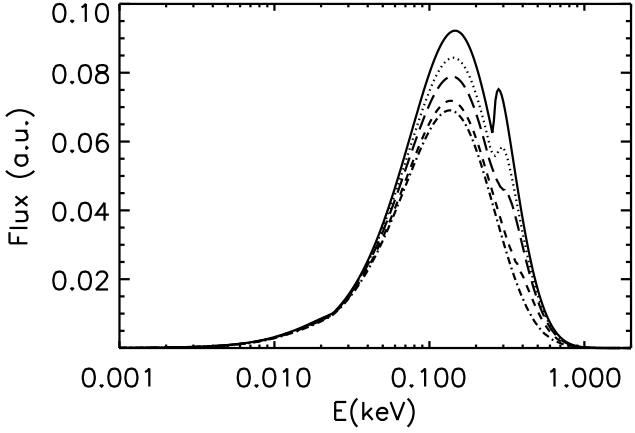}}
\caption{Phase dependent emitted spectrum (unabsorbed) of a FF model
with $B_{p} = 10^{13}$G and $T_{\rm c} = 5 \times 10^{7}$ K.
We have assumed ${\cal O} = \pi/2$ and ${\cal B} = \pi/2$.
From bottom to top, the different lines correspond to phase angles
of 0, 30, 45, 60, and 90$^\circ$. The hot polar cap (and therefore the magnetic
field axis) is pointing to the observer when the phase angle is 0$^\circ$.
}
\label{phase}
\end{figure}

Since most spectral fits to real data are made with simple blackbody models,
we have taken one of our models (FF, $B_{p} = 10^{13}$ G, $T_{\rm c} = 5 \times 10^{7}$ K)
and fitted our results to a single blackbody. We have assumed a column density
of $n_H=1.5\times 10^{20}$ cm$^{-2}$, typical of galactic interstellar medium
absorption. The comparison between this model, and a BB fit is shown in Fig. \ref{fit1}.
The X-ray part of the spectrum is well fitted by a single blackbody but our model
predicts an optical flux about a factor 4 larger than the blackbody fit to the high
energy part. This factor may vary depending on the magnetic field strength and geometry
and it is consistent with the systematic excess flux observed in the optical counterparts
of isolated neutron stars.
More interestingly, the condensed surface models also predict the
existence of an edge at an energy $E \approx \hbar \left(\omega_{B_i} + \omega_p^2/\omega_{B_e}\right)$
\citep{Lai05,paper1}, where $\omega_p = (4\pi e^2 n_e/m_e)^{1/2}$ is the electron plasma frequency,
that for typical magnetic fields ($10^{13}-10^{14}$ G) falls in the range 0.2-0.6 keV
(depending also on the gravitational redshift).
Some spectral features have been reported in that range although they are usually associated to proton
synchrotron lines. The only object for which an independent estimate of the magnetic field
is available is J0720, for which the measure of $\pdot = 6.98 \pm 0.02 \times 10^{-14} (s~s^{-1})$
implies $B=2.4 \times 10^{13}$ G \citep{KK05}. The observed spectral feature is fitted by
a Gaussian absorption line at an energy of 0.27 keV, and has been associated to
cyclotron resonance scattering of protons in a magnetic field with $B=5 \times 10^{13}$ G
\citep{Hab04b}. Assuming a magnetic field strength (from $\pdot$) of $B=2.4 \times 10^{13}$ G, the
condensed surface model predicts a phase dependent edge at an energy (local) of 0.35 keV, 
which would imply a redshift of $z=0.29$.
The phase dependent emitted spectrum for one of our models (FF) is shown in Fig. \ref{phase}.
We have taken ${\cal O}=\pi/2$ and ${\cal B} = \pi/2$. The feature is strongly dependent
on the orientation, being stronger when the magnetic field axis is pointing to the observer
and practically undetectable when the magnetic axis is normal to the direction of
observation. The angles have been chosen to show the most extreme case, where the 
variability is very large.
As reported in Table \ref{pfff}, this particular model has a maximum pulsed fraction of
0.24.

%%%%%%%%%%%%%%%%%%%%%%%%%%%%%%%%%%%%%%%%%%%%%%%%%%%%%%%%%%%%%%%%%%%%%%%%%%%%%%%%

%%%%%%%%%%%%%%%%%%%%%%%%%%%%%%%%%%%%%%%%%%%%%%%%%%%%%%%%%%%%%%%%%%%%%%%%%%%%%%%%

%%%%%%%%%%%%%%%%%%%%%%%%%%%%%%%%%%%%%%%%%%%%%%%%%%%%%%%%%%%%%%%%%%%%%%%%
\section{Conclusions.}

In this paper we have presented the results of detailed calculations of the
temperature distribution in the crust and condensed envelopes of neutron 
stars in the presence of strong magnetic fields. The surface
temperature distribution has been calculated by obtaining 2D stationary 
solutions of the heat diffusion equation with anisotropic thermal conductivities.
From the variety of strengths and configurations of the magnetic field explored,
we conclude that variations in the surface temperature of factors 2-10 are easily
obtained with magnetic fields in the range ($B\ge 10^{13}$-$10^{14}$ G). 
The average luminosity (or the inferred effective temperature) does not depend
much on the strength of the magnetic field, but it is drastically affected by
the geometry, in particular by the existence of a toroidal component.
The toroidal field acts effectively as a heat insulator forcing heat to 
flow towards the poles. Therefore, it is the particular geometry (a priori unknown)
of the magnetic field what eventually determines the size of the hot polar caps.
A back of the envelope calculation to estimate the size of the polar cap is the following.
For purely radial magnetic fields the non-magnetic solution is not affected while
for purely tangential fields the temperature gradient is quite larger. Therefore
the hot polar cap will be determined by the angular size of the region in which
the magnetic field is nearly radial. For a classical dipole, the condition $B_r^2=B_{\theta}^2$ 
leads to $\sin\theta \approx 2/\sqrt{5}$, which gives a hot polar cap of about
$63^\circ$. For FF models, for example, the condition $B_r^2=B_{\theta}^2+B_{\phi}^2$
leads to $\sin\theta \approx 2/\sqrt{5+\mu^2 R^2}$. The configurations we have employed
correspond to $\mu R \approx 16$, which gives an estimate of the size of the hot polar 
cap of $7^\circ$, in good agreement with the numerical results.

For purely poloidal configurations, the surface temperature is high in a large
fraction of the star surface and lower in a narrow equatorial band, while for
configurations with toroidal components of the same strength as the poloidal one
the temperature distributions is more close to hot polar cap with a large cooler
area at low latitudes. Thus, this latter family of models shows larger pulsation
amplitudes and optical excess flux, in very good
agreement with the observed properties of isolated neutron stars.  We defer to 
future work for detailed fitting of real data with our models, but preliminary
calculations show that the spectral energy distribution and its variability
can be easily explained without fine tunning of the model parameters.
This can be interpreted as indirect evidence of the existence of toroidal fields
in the crust and envelopes of NSs.
We have also investigated the influence of some relevant inputs such as the physical
conditions of the surface (condensed, gaseous) by varying the outer boundary
conditions, i.e., the emissivity at a given temperature and $\vec{B}$. We found
that the main conclusions remain qualitatively unchanged, although quantitative
differences can arise. We have also explored the effect of having different impurity
content, finding that their effect is not important in general, being only visible 
in models without toroidal components.

Another interesting result is that the condensed surface models predict the
existence of an edge at an energy $ E \approx \hbar \left(\omega_{B_i} + \omega_p^2/\omega_{B_e}\right)$
that for typical magnetic fields falls in the range 0.2-0.6 keV, where some
spectral features have been reported, and usually associated to proton
synchrotron lines. The energy of the spectral feature observed in J0720, as well
as its pulsation amplitude predicted by our models are consistent with 
the inferred magnetic field.
We also plan to extend our work to calculations with stronger magnetic fields and
higher temperatures, typical condition of magnetars (SGRs, AXPs). 
The mean caveat that we must point out is the large uncertainty in the particular
structure of the magnetic fields inside neutron stars and the need of a full
2D calculation of the relativistic structure of neutron stars with arbitrary magnetic fields.
The bottom line is that magnetic fields do change significantly the thermal
emission from isolated neutron stars and cannot be overlooked if one expects
to infer valuable information (radius, gravitational redshift, composition)
from the observed spectral energy distribution. What one infers from the blackbody
fits to X-ray observations (assuming a known distance to the object) 
is the product $T_{\infty}^4 R_\infty^2$,
and model dependent variations in the estimation of the effective temperature 
translate into the estimate of the radius. 
Our models with toroidal components result in inferred radii about
a factor 3-5 larger than the BB radius or than the inferred radius from a model with 
only poloidal component. This naturally solves the problem of the apparent smallness
of some isolated neutron stars.

If the existence of strong magnetic fields in isolated NSs is confirmed, 
we will need more detailed calculations coupling the evolution of
the magnetic field with the temperature before we can establish
firm constraints on NS properties by fitting observational data.

%\bigskip
\begin{acknowledgements}
We thank to D. Yakovlev, Dany Page, and Ulrich Geppert for very 
useful comments on the manuscript.
This work has been supported by the Spanish Ministerio
de Ciencia y Tecnolog\'{\i}a grant AYA 2004-08067-C03-02.
JAP is supported by a {\it Ram\'on y Cajal} contract from the Spanish MCyT.
\end{acknowledgements}

%%%%%%%%%%%%%%%%%%%%%%%%%%%%%%%%%%%%%%%%%%%%%%%%%%%%%%%%%%%%%%%%%%%%%%%%

\begin{table*}
\caption{Coefficients of the fit to the emissivity from a condensed surface
(Eq. \ref{alfafit}) for $B=5 \times 10^{13}$ G taking into account the effect of the
motion of ions.}
\begin{tabular}{lcccccc}
%\noalign{\smallskip}
\hline
\hline
\noalign{\smallskip}
 $a_{i,j}$ &  1     &     2     &     3     &   4       &    5      &   6     \\
\hline\noalign{\smallskip} 
     1  & 0.451428  & 0.652402  & -0.329039 & 0.123172  & 0.107814  & -0.0967540   \\
     2  & -0.304007 & -0.626414 & -5.79256  & 17.3578   & -20.8844  & 9.57023  \\
     3  & -0.378297 & 4.48855   &  3.42040  & -26.5822  & 38.1295   & -18.8900 \\
     4  & 0.562032  & -3.96036  & -0.866190 & 17.5741   & -26.3781  & 13.3072  \\
     5  & -0.226386 & 1.36369   & 0.0520957 & -5.29775  & 8.03772   & -4.06952 \\
     6  & 0.0299993 & -0.169923 & 0.0232908 & 0.568170  & -0.880843 & 0.450272 \\ 

\hline\noalign{\smallskip}
\end{tabular}
\label{ion513tab}
\end{table*}

\begin{table*}
\caption{Same as table \ref{ion513tab} without taking into account the effect of the motion
of ions.}
\begin{tabular}{lcccccc}
%\noalign{\smallskip}
\hline
\hline
\noalign{\smallskip}
 $a_{i,j}$ &  1            &      2       &     3        &   4          &    5          & 6            \\
\hline\noalign{\smallskip}
     1     & 0.0520960   & 0.906625 & -2.27950 & 3.50014 & -2.73453  & 0.831342   \\
     2     & -0.092588   & 3.96721  & -6.00109 & 1.76058 & 3.73908   & -2.65634  \\
     3     & 0.1884173   & -6.88457 & 16.2169  & -20.2954 & 13.1221  & -3.36603 \\
     4     & -0.0905443  & 5.4989   & -15.7136 & 24.2137  & -19.2636 & 6.14122   \\
     5     & 0.0197174   & -1.92340 & 6.01936 & -9.94622 & 8.30053  & -2.73754  \\
     6     & -0.001500  & 0.239160  & -0.782216 & 1.33117 & -1.13048 & 0.376769  \\

\hline\noalign{\smallskip}
\end{tabular}
\label{wion513tab}
\end{table*}

\begin{table*}
\caption{Same as table \ref{ion513tab} with $B=10^{13}$ G.}
\begin{tabular}{lcccccc}
\hline
\hline
\noalign{\smallskip}
 $a_{i,j}$ &  1            &      2       &     3        &   4          &    5          & 6            \\
\hline\noalign{\smallskip}
     1     & 0.37727   & -0.0297095 & 0.172343  & 0.934048  & -1.86330 & 1.05495      \\
     2     & -1.32853  & 7.41110  & -16.7803 & 15.8174  & -4.53211  & -1.57610  \\
     3     & 1.70549   & -7.45357 & 16.4152  & -14.9966 & 5.40906   & 0.726946  \\
     4     & -0.826587 & 1.60020  & 2.15163  & -13.0149 & 14.5113   & -5.62832 \\
     5     & 0.185560  & 0.229112  & -3.75523 & 9.92179  & -9.52118  & 3.30446  \\
     6     & -0.0162908 & -0.076025 & 0.652672  & -1.58083 & 1.48421   & -0.505131 \\
\hline\noalign{\smallskip}
\end{tabular}
\label{ion13tab}
\end{table*}

\begin{table*}
\caption{Same as table \ref{wion513tab} with $B=10^{13}$ G.}
\begin{tabular}{lcccccc}
\hline
\hline
\noalign{\smallskip}
 $a_{i,j}$ &  1            &      2       &     3        &   4          &    5          & 6            \\
\hline\noalign{\smallskip}
     1     &  0.0682268 & 0.945109 & -2.14347 & 2.99701  & -2.20175  & 0.647961     \\
     2     & -0.269514  & 4.23065  & -6.58908 & 2.17098  & 4.27706   & -3.28306 \\
     3     & 0.323785   & -3.69871 & 2.05278  & 6.96692  & -11.4661  & 5.26209  \\
     4     & 0.030067   & -0.483140 & 11.2452  & -27.8763 & 26.7549   & -9.23072 \\
     5     & -0.066218  & 0.77907  & -6.42987 & 14.4756  & -13.4195  & 4.50075  \\
     6     & 0.0118741  & -0.131711 & 0.949772  & -2.10127 & 1.94095   & -0.648770 \\

\hline\noalign{\smallskip}
\end{tabular}
\label{wion13tab}
\end{table*}

\end{document}